\documentclass[showpacs, aps, pre, preprint, superscriptaddress, amssymb]{revtex4}

\usepackage{graphicx}
\usepackage{color}
\usepackage{amsmath}
\usepackage{pdfsync}
\begin{document}

%\preprint{}

\title{Nonequilibrium steady solutions of the Boltzmann equation}
%\title{Nonequilibrium steady states in the homogenous isotropic Boltzmann equation}
%\title{Turbulent cascades in the \blue{isotropic} \red{[PIETRO: maybe it is better to be precise]} Boltzmann kinetic equation}



\author{Davide Proment}
\email{davideproment@gmail.com}
\homepage{http://www.to.infn.it/~proment}
\affiliation{Dipartimento di Fisica Generale, Universit\`{a} di Torino, Via Pietro Giuria 1, 10125 Torino, Italy}
\affiliation{INFN, Sezione di Torino, Via Pietro Giuria 1, 10125 Torino, Italy}

\author{Miguel Onorato}
\affiliation{Dipartimento di Fisica Generale, Universit\`{a} di Torino, Via Pietro Giuria 1, 10125 Torino, Italy}

\author{Pietro Asinari}
%\homepage{http://www.institution.edu/~Author}
\affiliation{Dipartimento di Energetica, Politecnico di Torino, Corso Duca degli Abruzzi 24, 10129 Torino, Italy}

\author{Sergey Nazarenko}
\affiliation{Mathematics Institute, The University of Warwick, Coventry, CV4-7AL, UK}

%\thanks{}
%\affiliation{}


\date{\today}

\begin{abstract}
We report a study of the homogeneous isotropic Boltzmann equation for an open system. 
We seek for nonequilibrium steady solutions in presence of forcing and dissipation. 
Using the language of weak turbulence theory, we analyze the possibility to observe Kolmogorov-Zakharov steady distributions. 
We derive a differential approximation model and we find that the expected nonequilibrium steady solutions have always the form of warm cascades. 
We propose an analytical prediction for relation between the forcing and dissipation and the thermodynamic quantities of the system. 
Specifically, we find that the temperature of the system is independent of the forcing amplitude and determined only by the forcing and dissipation scales. Finally, we perform direct numerical simulations of the Boltzmann equation finding consistent results with our theoretical predictions.
\end{abstract}


\maketitle

% \section{}
% \subsection{}
% \subsubsection{}

\section{Introduction}

Systems  in a steady state are characterized by observables  that do not  change in time; they can be either in equilibrium
or out of equilibrium. Systems in nonequilibrium
steady states have net currents (fluxes): 
examples of nonequilibrium steady-state
systems include an object in contact with two thermal sources at different
temperatures, for which the current is a heat flux; a resistor with electric
current flowing across it; the kinesin-microtubule system, for which
kinesin motion is the current. Most biological systems, including molecular machines and even whole cells, are in nonequilibrium states \cite{bustamante:43}. In particular, biological systems rely on
a continuous flux of energy and/or particles supplied by some proper environmental
reservoirs.

In statistical mechanics, investigating the general properties of a system in contact
with reservoirs, namely an open system, is a long lasting problem (e.g. see the second problem
discussed by E.H. Lieb on the occasion of the award of the Boltzmann medal \cite{Lieb1999491}),
even though these theoretical challenges are sometimes neglected in applied engineering at
large. The difficulties arise from the fact that finding the large deviation functional for a
stationary state with fluxes is still an open problem (see \cite{PhysRevLett.100.230602} and
references therein). In the present work, for focusing our attention and considering an affordable
goal, we consider the kinetic theory of gases. In particular, we consider a
system composed of a large number of interacting particles, comparable to the Avogadro number.
The Boltzmann kinetic equation (BKE) describes the time evolution
of the single-particle distribution function, which provides a statistical description of
the positions and velocities (momenta) of the gas molecules.
This integro-differential kinetic equation, proposed by Boltzmann at the end of the XIX century, has been derived starting from the phase-space Liouville equation, assuming the {\it stosszahl ansatz} \cite{cercignani1988}. Its equilibrium state, which maximizes the entropy measure, is the Maxwell-Boltzmann distribution. 
In case of small deviations from the local equilibrium, it is possible to systematically derive hydrodynamic equations for macroscopic quantities of the system; e.g., in the lowest order approximation for small departures from equilibrium, the Navier-Stokes equations \cite{cercignani1988}.

%In the hypothesis of an homogenous system the transport term of the Boltzmann equation is null and the evolution is completely ruled by the collision integral which mimics all possible binary collisions throw particles. This is a particular case of what is called in the literature kinetic equation: in the following we will then use the acronym Boltzmann kinetic equation (BKE).

Kinetic equations have also been studied in the framework of wave turbulence theory \cite{zakharov41kst} where it has been shown that other solutions with respect to thermodynamic solutions can be stationary states of the system, in case of external forcing and dissipation. These distributions, which have usually the form of power-laws in momentum space, are called Kolmogorov-Zakharov (KZ) and they represent constant flux of conserved quantities similar to the Kolmogorov  energy cascade in strong Navier-Stokes turbulence \cite{kolmogorov41, frisch1995t}. These solutions, named cascade solutions, become important when considering an open system, i.e. with forcing and dissipation terms. They have been studied for a great variety of  weakly nonlinear dispersive models: examples can be found in water waves \cite{janssen2004iow, onorato2002fdw, dyachenko2003wtg}, internal waves \cite{lvov2004eso},  nonlinear optics \cite{dyachenko:1992hc}, Bose-Einstein condensation \cite{berloff2002ssn, nazarenko2006wta, proment:051603}, magnetohydrodynamics \cite{galtier2000wtt}.

An out of equilibrium description of the Boltzmann equation using the KZ solutions was first devised in \cite{kats1975} considering different types of interaction potential between particles. Problems of interaction locality scale-by-scale and wrong flux direction were pointed out.
In particular in \cite{kats1976} Kats showed that for all realistic physical situations the direction of the cascades in the system is always in the wrong orientation with respect to the one predicted by the Fj{\o}rtoft theorem
\footnote{This theorem, originally put forward by Fj{\o}rtoft in 1953 for the 2D turbulence, says that that
the integral whose density grows fastest with the wavenumber/momentum must cascade from low to high wavenumbers/momenta. The other integral must cascade inversely, from high to low wavenumbers/momenta. For the classical particles, this means that the energy flux must be from low to high momenta, and the flux of particles must  be toward low momenta; see Section \ref{ssec:fluxKZ}.}.
When a  formal KZ solution has a flux direction contradicting with  the Fj{\o}rtoft theorem, this spectrum (even if local) cannot be established because it cannot be matched to any physical forcing and dissipation at the ends of the inertial range.
For example in \cite{dyachenko:1992hc}, the particle cascade KZ solution was found to be of this type in the two-dimensional nonlinear Schr\"{o}dinger equation model the authors argued that in this case the KZ solution is not achievable and a mixed state, with both a cascade and a thermodynamic components were proposed.
 %model this fact happens the cascades are usually carry on by the thermodynamic solutions: they represent small deviations to the equilibrium and they are called warm cascades. This has been observed in some weak nonlinear dispersive systems for some space dimensions \cite{dyachenko:1992hc} or i
Another example of mixed cascade-thermodynamic states can be found in the context of three-dimensional Navier-Stokes turbulence \cite{nazarenko2004warm}, where such mixed  states were called {\it warm cascades}
\footnote{In Navier-Stokes the warm cascades correspond to so called bottleneck phenomenon which arises in numerics due to an energy flux stagnation near the maximum wave-number.}.

The present manuscript will focus on warm cascades found in the homogenous isotropic Boltzmann equation (HIBE) and in particular it will answer to the following important questions.
\begin{itemize}
\item What is precisely the relation between the conserved quantity fluxes and the thermodynamics quantities of the system?
\item How does this relation depends on the forcing and dissipation rates and acting scales?
\end{itemize}
To answer the above questions we will perform numerical simulations of the homogeneous isotropic Boltzmann equation with forcing and dissipation.  
We will then use a diffusion approximation model (DAM) to derive analytical predictions on how the thermodynamic quantities, temperature and chemical potential, are related to fluxes, forcing and dissipative scales. We will then test these predictions by numerically simulating both DAM and the complete homogenous isotropic Boltzmann equation.

The work is organized as follows: in Section \ref{HIBE} we review the properties of the Boltzmann equation for the homogeneous isotropic case; in Section \ref{DAM} we introduce DAM  and we derive the analytical predictions; Section \ref{results} is dedicated to numerical results of DAM and HIBE; in Section \ref{conclusions} we draw the conclusions. A set of Appendixes also provide detailed calculations of those results which are briefly reported in the main text.

\section{The Boltzmann kinetic equation\label{HIBE}}
The Boltzmann kinetic equation describes the time evolution of the single-particle distribution function, which provides a statistical description for the positions and momenta of the gas molecules: the function $ n(\mathbf{x}, \mathbf{k}, t) $ express a probability density function in the one-particle phase space $ \mathbb{R}_{\mathbf{x}}^d \times \mathbb{R}_{\mathbf{k}}^d $ with respect to time, where $ d $ is the dimension.
Note that we denote the momentum variable with the letter $ \mathbf{k} $ instead of the conventional $ \mathbf{p} $ to follow the common notation of wave turbulence \cite{zakharov41kst}.
The Boltzmann equation takes the following form:
\begin{equation}
\frac{\partial n}{\partial t} (\mathbf{x}, \mathbf{k}_1, t) + \frac{\mathbf{k}_1}{m} \cdot \frac{\partial n}{\partial \mathbf{x}} (\mathbf{x}, \mathbf{k}_1, t) = I_{coll} (\mathbf{x}, \mathbf{k}_1, t),
\label{eq:BE}
\end{equation}
%\red{[DAVIDE: check dimensionality!]} 
where
\begin{equation}
I_{coll}=\int_{-\infty}^{+\infty} W_{12}^{34} \left[n(\mathbf{x}, \mathbf{k}_3, t)n(\mathbf{x}, \mathbf{k}_4, t)-n(\mathbf{x}, \mathbf{k}_1, t)n(\mathbf{x}, \mathbf{k}_2, t)\right] d\mathbf{k}_2 d\mathbf{k}_3 d\mathbf{k}_4
\end{equation}
sums the effect of the two-body collisions of particles with all possible values of momenta. The form of the collision integral we are reporting is equivalent to the standard one and corresponds to Eq. (4.18), page 64 in Cercignani's book \cite{cercignani1988}.
Here  $ W $ describes synthetically the scattering amplitude transition $ 2 \rightarrow 2 $ as a function  of the momenta of the interacting particles. As we consider elastic collisions, the  general way to express $ W $ is
\begin{equation}
W_{12}^{34}=\Gamma_{12}^{34} \delta(\mathbf{k}_1+\mathbf{k}_2-\mathbf{k}_3-\mathbf{k}_4) \delta(|\mathbf{k}_1|^2+|\mathbf{k}_2|^2-|\mathbf{k}_3|^2-|\mathbf{k}_4|^2),
\end{equation}
where $ \delta $-functions assure conservation of the total momentum and the total kinetic energy (which is proportional to $ |\mathbf{k}|^2 $) of incoming and outgoing particles. The collision probability, expressed by $ \Gamma_{12}^{34} \equiv\Gamma(\mathbf{k}_1, \mathbf{k}_2 | \mathbf{k}_3, \mathbf{k}_4) \ge 0 $, is invariant under permutations $ \{1, 2\}\rightarrow\{2, 1\} $,  $ \{3, 4\}\rightarrow\{4, 3\} $, and $ \{1, 2\}\rightarrow\{3, 4\} $.
In the present paper we will consider the case of three-dimensional rigid spheres with diameters $ \sigma $ and mass $ m $, for which $ \Gamma $ simply results in $ \Gamma_{12}^{34} =2\sigma^2/m $ \cite{cercignani1988}. For other interaction potentials, as Coulomb or Born approximation, refer to \cite{karas1976, kats1976}.

For the purposes of our work, we consider a homogeneous and isotropic (in physical space $ \mathbb{R}_{\mathbf{x}}^d $) system with the one-particle probability density function independent of  $ \mathbf{x} $  and its momentum dependency coming only via  the modulus $ k=|\mathbf{k}| $, so $ n(\mathbf{x}, \mathbf{k}, t) \rightarrow n(k, t) $. It is useful  to express the distributions  in the energy space $ \omega_i = |\mathbf{k}_i|^2 $ where we use again the notation $ \omega $ for the energy in analogy with wave turbulence. Then, the particle density in $\omega$-space satisfies the relation $ \int N(\omega, t) d\omega = \int n(k, t) d\mathbf{k} $ or, in the other words, $ N(\omega, t)=n(\omega, t)\,\Omega\,\omega^{\frac{d-1}{2}} \left|\frac{dk}{d\omega}\right| $, where $ \Omega $ is the solid angle. 
After these considerations Boltzmann equation (\ref{eq:BE}) simplifies to the homogeneous isotropic Boltzmann equation (HIBE):
\begin{equation}
\frac{\partial N_1}{\partial t}=\int_0^{\infty} S_{12}^{34} (n_3n_4-n_1n_2)\delta(\omega_1+\omega_2-\omega_3-\omega_4) d\omega_{2}d\omega_{3}d\omega_{4},
\label{eq:HIBE}
\end{equation}
where we denote for brevity $ N_i=N(\omega_i, t) $ and $ n_i=n(\omega_i, t) $, and the functional
\begin{equation}
S_{12}^{34} = (\omega_1\omega_2\omega_3\omega_4)^{\frac{d-1}{2}} \left|\frac{dk_1}{d\omega_1}\right| \left|\frac{dk_2}{d\omega_2}\right| \left|\frac{dk_3}{d\omega_3}\right| \left|\frac{dk_4}{d\omega_4}\right| \int_{\Omega} \Gamma_{12}^{34} \delta(\mathbf{k}_1+\mathbf{k}_2-\mathbf{k}_3-\mathbf{k}_4)  d\Omega_1 d\Omega_2 d\Omega_3 d\Omega_4
\end{equation}
takes into account the change of coordinates and the average over solid angles.
Hereafter, we always consider a three-dimensional gas of hard-sphere particles in a non-dimensional form with $ m=1 $ and $ \sigma^2=8 $ .
Then the functional simply results in $ S_{12}^{34}=2\pi \min\left[\sqrt{\omega_1}, \sqrt{\omega_2}, \sqrt{\omega_3}, \sqrt{\omega_4}\right] $ (see Appendix
%\ref{A:HIBE-hs} and
\ref{delta3d} for details of the angular integration).
%\begin{equation}
%S_{12}^{34}=2\pi \min\left[\sqrt{\omega_1}, \sqrt{\omega_2}, \sqrt{\omega_3}, \sqrt{\omega_4}\right].
%\label{eq:HIBE-hs}
%\end{equation}

The HIBE has two conserved quantities, the mass and energy densities,
\begin{equation}
\begin{split}
& \rho_M=\int_{-\infty}^{+\infty} n(\omega, t) d\mathbf{k} = 2\pi \int_0^{+\infty} n(\omega, t) \sqrt{\omega} d\omega , \\
& \rho_E=\int_{-\infty}^{+\infty} n(\omega, t) {k}^2 d\mathbf{k} = 2\pi \int_0^{+\infty} n(\omega, t) \omega^{\frac{3}{2}} d\omega .
\end{split}
\label{eq:densities}
\end{equation}
Note that $ \rho_M $ and $ \rho_E $ are always constant in time for any distribution $ n $ and interaction potential, due to the fact that collisions are $ 2\rightarrow 2 $ and elastic. This is evident by evaluating their time derivatives using equation (\ref{eq:HIBE}): the symmetries with respect to the integration indices immediately show that these quantities are zero.

\subsection{Steady solutions}

\subsubsection{Equilibrium in a closed system}
The HIBE (\ref{eq:HIBE}) is an integro-differential equation with no general analytic solution. It is easy, however, to look for  steady (time independent) solutions. In closed system, i.e. without forcing and/or dissipation mechanisms, the only steady solution corresponds to the thermodynamic equilibrium described by the Maxwell-Boltzmann (MB) distribution,
\begin{equation}
n_{MB}(\omega)= e^{-\frac{\omega+\mu}{T}}= A e^{-\frac{\omega}{T}},
\label{eq:MB}
\end{equation}
where $ A=e^{-\frac{\mu}{T}} $ and constants $\mu$ and $T$ have the meaning of the chemical potential and the temperature respectively (we consider the natural unit system, where the Boltzmann constant is one). Validation is trivial by plugging (\ref{eq:MB}) into (\ref{eq:HIBE}): for any value of $ T $,  $ \mu $ and the interaction potential $ S_{12}^{34} $, the $ \delta $-function assures that the integrand is zero.
%\red{Of course, once we have a well-defined the equilibrium state, the expectation value of macroscopic quantities can be easily evaluated. [PIETRO: NO! It is the other way around: $n_{MB}$ is a function of $n$. One computes the moments of $n$ first and then one uses these values for computing the local equilibrim. Delete this sentence]}
Moreover, the total mass density of the system is $ \rho_M = A \left(\pi T\right)^{\frac{3}{2}} $,  the total energy density is $ \rho_E = \frac{3}{2}A\pi^{\frac{3}{2}} T^{\frac{5}{2}} $, and any other moment of $ \omega $, due to the bi-parametric nature of the MB distribution, is a function of $ \rho_M$ and $ \rho_E$. The H theorem states that in a closed system any out of equilibrium distribution with defined mass and energy densities will always relax to the MB distribution having same $ \rho_M $ and $ \rho_E $.  

\begin{figure}
\includegraphics[scale=1.6]{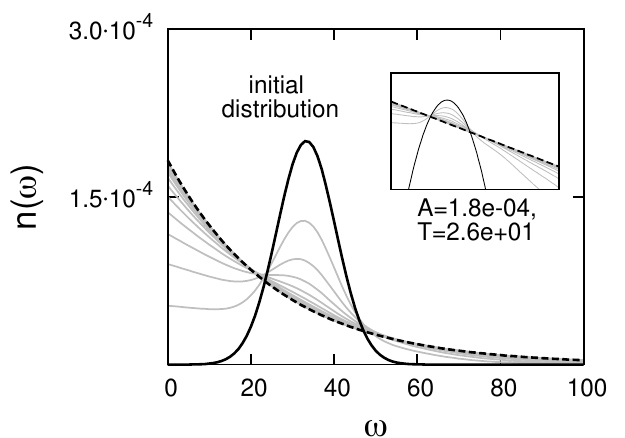}
\caption{Numerical computation of  HIBE with an initial Gaussian shaped distribution (continuos black line): intermediate states are shown with gray lines and final steady distribution with dashed black line. The latter has a MB behavior (\ref{eq:MB}) with fitted parameters $ A $ and $ T $ printed in figure. The inset shows the same plot in lin-log scale.
\label{fig:relax}}
\end{figure}

In Fig. \ref{fig:relax} we show a numerical simulation of the HIBE  with initial condition given by a Gaussian function centered around a particular value of energy;  as it is clear from the figure, the 
initial condition relaxes to the MB distribution. The numerical algorithm used to perform this simple example will be discussed in Section \ref{hibeCode}.
We can observe that the initial condition evolves reaching an equilibrium MB distribution: the exponential behavior become evident by observing the inset where we plot it lin-log plot scale. Moreover by fitting the results with the MB function we can find the thermodynamic quantities $ A $ and $ T $: those correspond exactly to ones expected knowing initial mass and energy densities (note that now integrals (\ref{eq:densities}) are evaluated from 0 to a finite value of $ \omega $ due to numerical finiteness of $ \omega $-space).

\subsubsection{Nonequilibrium steady states}
Now, what can we expect in an open system driven by external forcing and dissipation mechanisms? 
We will answer this question keeping in mind the main results of the wave turbulence theory.
Part of this theory is dedicated to study steady solutions to kinetic equations in the power-law form, $ n(\omega) \sim \omega^{-x} $, where the constant $ x $ assumes different values depending on the considered wave system.
%for which (\ref{eq:HIBE}) is an example and particular attention is focused on steady solutions.
It is sometimes possible to find the so-called Kolmogorov-Zakharov (KZ) solutions $ n_{\small  KZ} (\omega) \sim \omega^{-x} $ which correspond to constant fluxes of conserved quantities through scales. The KZ distribution always appears in a range of scales, known as inertial range, between the forcing and dissipation were the source and sink are located.

As already mentioned, the HIBE conserves the number of particles and the energy, and so one could expect to observe two turbulent KZ cascades. The KZ exponent $ x $ can be evaluated by applying the standard Zakharov transformations \cite{zakharov41kst}, by dimensional analysis \cite{connaughton:2003lg}, or by using the method (equivalent to Zakharov transformation) proposed by Balk \cite{balk2000kzs}.
We have chosen the last one and the complete analytical calculations are presented in Appendix \ref{A:KZ}.
The KZ exponents depend on the scaling behavior of the scattering term $ \Gamma_{12}^{34} $ and on the dimension $ d $ of the particle system. For the particular case of three-dimensional hard spheres we have
\begin{equation}
\begin{split}
& \mbox{\underline{constant particle flux $ \eta $} } \ \ \  \Longrightarrow \ \ \  n_{KZ}(\omega)\sim\omega^{-\frac{7}{4}}, \\
& \mbox{\underline{constant energy flux $ \epsilon $}} \ \ \  \Longrightarrow \ \ \  n_{KZ}(\omega)\sim\omega^{-\frac{9}{4}}.
\end{split}
\label{eq:KZ}
\end{equation}

The simplest way to mimic an open system where steady nonequilibrium distributions of the form of turbulent KZ solutions can be establish is to consider a forced-dissipated HIBE
\begin{equation}
\frac{\partial N}{\partial t}(\omega_1, t)= I_{coll}(\omega_1, t) + F(\omega_1) - D(\omega_1)N(\omega_1).
\label{eq:fdHIBE}
\end{equation}
The forcing $ F $ is constant in time and very narrow near a particular energy value $ \omega_f $: with this choice the incoming fluxes of particles $ \eta $ and energy $ \epsilon $ roughly satisfy relation $ \epsilon=\omega_f\, \eta $.
The dissipation term $ D $ is implemented as a filter which removes,  at each iteration time, energy and particles outside of the domain $ \omega \in \left(\omega_{\min}, \omega_{\max} \right) $. Further details on the numerical scheme are explained in Section \ref{hibeCode}. What happens if we try to solve numerically such forced/damped integro-differential equation? 

\begin{figure}
\includegraphics[scale=1.6]{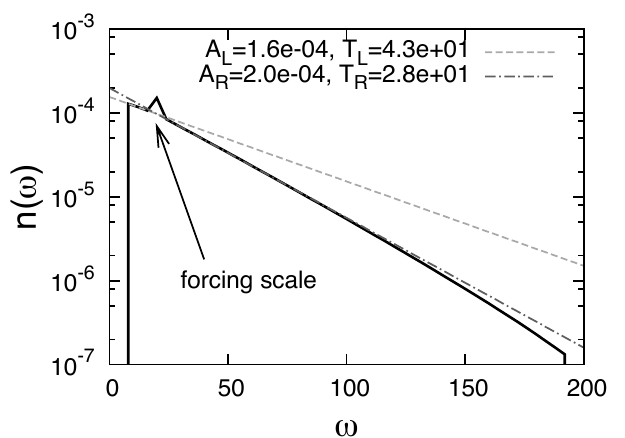}
\caption{An example of HIBE (\ref{eq:fdHIBE}) steady state shown with black line in lin-log scale. Simulation parameters are: $ \omega_{\min}=5 $, $ \omega_f=22 $, $ \omega_{\max}=195$ and $ F=10^{-5} $, and $ \omega_{cutoff}=200$. The dashed and point/dashed lines are left and right branch best fits obtained with the MB distribution (\ref{eq:MB}): fitting parameters are reported in label.
\label{fig:exampleEC}}
\end{figure}

\begin{figure}
\includegraphics[scale=1.6]{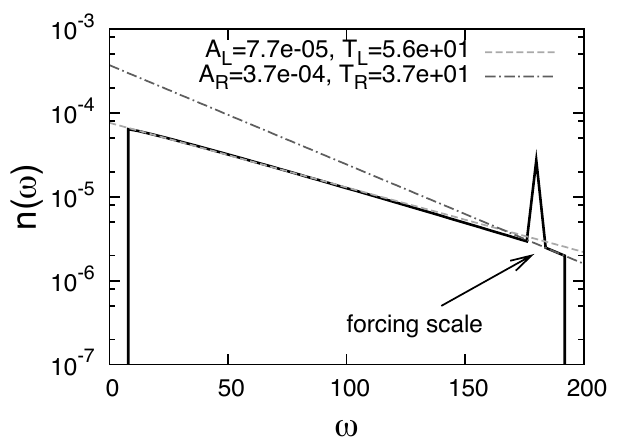}
\caption{An example of HIBE (\ref{eq:fdHIBE}) steady state shown with black line in lin-log scale. Simulation parameters are: $ \omega_{\min}=5 $, $ \omega_f=182 $, $ \omega_{\max}=195$ and $ F=10^{-5} $, and $ \omega_{cutoff}=200$. The dashed and point/dashed lines are left and right branch best fits obtained with the MB distribution (\ref{eq:MB}): fitting parameters are reported in label.
\label{fig:examplePC}}
\end{figure}

In Fig. \ref{fig:exampleEC} and Fig. \ref{fig:examplePC} we plot the nonequilibrium steady states obtained with numerical simulations of the HIBE with forcing and dissipation; the initial conditions are characterized by $n(\omega,t=0)=0$ . The parameters in the simulations are $ \omega_{\min}=5 $, $ \omega_{max}=195 $, the forcing rate $ F=10^{-5} $. 
In Fig. \ref{fig:exampleEC}  forcing is located at $ \omega_{f}=22 $
and in Fig. \ref{fig:examplePC}  
at  $ \omega_{f}=182 $.
No power-law distributions, and so no KZ solutions (\ref{eq:KZ}), are observed (note that both plots are in lin-log scales), but instead one can see weakly perturbed exponential curves. 
We can attempt to measure the quantities 
$ T $ and $ A $ in (\ref{eq:MB}) by fitting our numerical curves;
however, those are not perfect straight lines (in the lin-log plot) and left and right branches with respect to forcing scale may give different results. 
For such reason we will denote by $ (\cdot)_L $ the quantities evaluate on the left brach and with $ (\cdot)_R $ the right ones. 
%and conjecture that 
%We note that measured temperature is always greater in the left zone while $ A $ is bigger to the right.
%taking wider intervals between the forcing and the dissipation scales would result in more similar left and right values.

\begin{figure}
\includegraphics[scale=1.6]{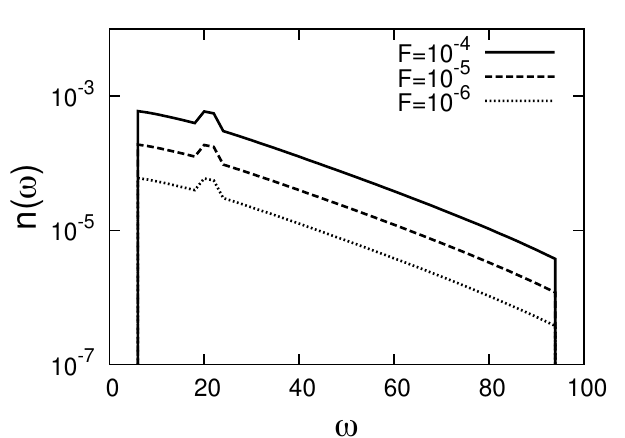}
\caption{Steady states of HIBE (\ref{eq:fdHIBE}) in lin-log scale obtained for different values of the forcing rate $ F $. Parameters are $ \omega_{cutoff}=200$, $ \omega_{\min}=5 $, $ \omega_f=21 $ and $ \omega_{\max}=95 $.  \label{fig:ke-flux}}
\end{figure}

Another example we analyze is the case where we fix the forcing and dissipative scales and change the forcing rate. Numerical results for final steady states evaluated for three different forcing amplitudes,  $F=10^{-4}, 10^{-5}, 10^{-6}$, are presented in Fig. \ref{fig:ke-flux}.
\begin{figure}
\includegraphics[scale=1.6]{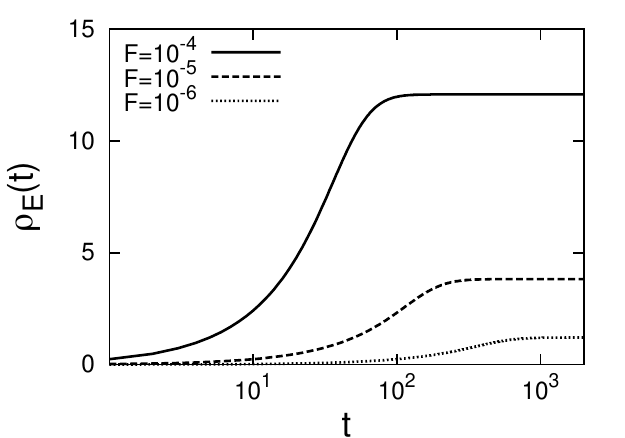}
\caption{Total energy densities $ \rho_E(t) $ in function of time for different forcing rates $ F $. For the system parameters refer to ones in Fig. \ref{fig:ke-flux}. \label{fig:spectra-en}}
\end{figure}
The effect of increasing the amplitude $F$ results in an upward shift of the curves. 
%: the higher is the particle flux $ \eta $, the higher is the amplitude of the spectrum. The second qualitative remark is that 
Therefore, qualitatively, the temperature appears to be the same for each value of the flux. The only difference is the speed at which the system, initially empty, reaches its steady state. In fig. \ref{fig:spectra-en} we show the energy density evolution (same line styles corresponds to same systems). 
%Note that  the parameters $ T $ and $ A $ of the MB distribution (\ref{eq:MB}) are not well defined because the curves are not perfect straight lines (in the linear-logarithmic plot) due to the insufficiently large inertial ranges, so that the flux corrections become substantial.

After these preliminary numerical results, a lot of questions can be posed. Why no KZ constant flux solutions are observed but just small deviations from MB distributions? What happens when forcing or dissipation scales are changed?  What is in general the relation between physical quantities such as  fluxes, forcing and dissipation scales and the MB parameters? The aim of this manuscript is to provide explanations to such phenomena and answer these questions.

\subsection{Locality of interactions}

For the KZ spectra to be valid mathematical (and therefore physically relevant) solutions, it is necessary that they  satisfy the locality condition. A spectrum is local when the collision integral converges. In other words, non-locality means that the collision integral is not weighted scale by scale but most of the contributions come from the limits of integration corresponding to the ends of the inertial range. Physically, the
 non-locality is in contradiction with the assumption that the flux of the relevant conserved quantity in the inertial range is carried only by the nearest scales.
Mathematically, locality guaranties that the KZ spectrum is a valid solution in an infinite inertial range, which is not
guarantied {\it a priori} because Zakharov transformation is not an identity transformation and could, therefore, lead to spurious solutions.

For the HIBE case, locality depends on the particular interaction potential, which affects the scaling of $ \Gamma_{12}^{34} $, and on the dimensionality of the system - for detailed calculations see Appendix \ref{A:KZ}.
Locality is not always found for both KZ solutions: for example for the Coulomb potential only the energy cascade is local, as shown in \cite{kats1975}. In the case of three-dimensional hard spheres considered in the present work, the criterion of locality is never satisfied for any of the two KZ solutions, which means that these solutions are un-physical and irrelevant in this model.

\subsection{The flux directions}
\label{ssec:fluxKZ}

Besides locality, another important requirement for establishment of the  KZ spectra is  the correctness of the flux directions for the respective conserved quantities. In a system where two quantities are conserved, the following Fj{\o}rtoft-type argument is used to establish which quantity must have a direct or an inverse cascade.

\subsubsection{The Fj{\o}rtoft argument}

Consider an open system where forcing scale $ \omega_f $ is widely separated from a low-$\omega$ dissipation scale $ \omega_{\min} $ and a high-$\omega$ dissipation frequency $ \omega_{\max} $, thus $ \omega_{\min} \ll \omega_f \ll \omega_{\max} $.
Because the energy density in the $\omega$-space is different from the particle density by factor $\omega$, the  forcing rate of the energy $\epsilon$ is related with the forcing rate of the particles $\eta$ as $\epsilon \sim \omega_f \eta$. Suppose that some energy is dissipated at the low scale $ \omega_{\min} $ at a rate comparable with the forcing rate $\epsilon$.
But then the particles would have to be dissipated at this scale at the rate proportional to $ \epsilon/ \omega_{\min} \sim \eta \, \omega_f/\omega_{\min} \gg \eta$, which is impossible in steady state because the dissipation cannot exceed the forcing. Thus we conclude that in the steady state the energy must dissipate only at $ \omega_{\max} $.
By a symmetric contradiction argument one can easily show that the only place where the particles can be dissipated in such systems is $ \omega_{\min} $.
This means that energy must have a direct cascade (positive flux direction) and particles an inverse cascade (negative flux direction).

\subsubsection{Flux directions in the HIBE}

\begin{figure}
\includegraphics[scale=1.6]{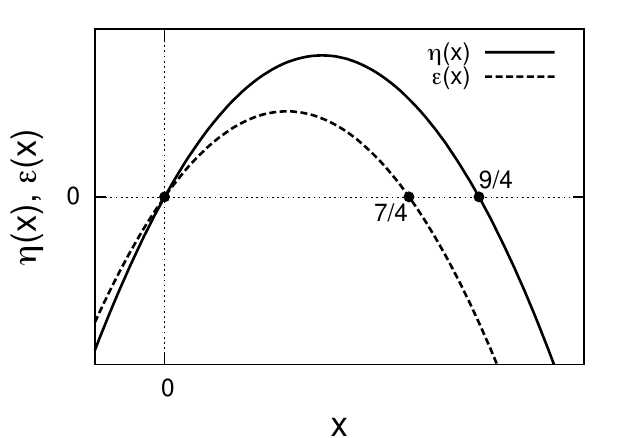}
\caption{Energy and particle fluxes on power-law solutions $ n(\omega, t)\sim\omega^{-x} $ as functions of $ x $ for the three-dimensional hard sphere model.
\label{fig:KZ-fluxes}}
\end{figure}

It has been proved in \cite{kats1976}, see also Appendix \ref{A:KZ} for details, that fluxes of the KZ solutions for all types of the interaction coefficient $ \Gamma $ have always the wrong directions with respect to the Fj{\o}rtoft argument requirements (in the case $ x > 0 $).
An alternative  way for finding the sign of the fluxes is considering them for general (not necessarily steady) power-law spectra $ n(\omega, t) \sim \omega^{-x} $ and plotting them as functions of $ x $ for a fixed $ (\omega, t) $, see Fig.  \ref{fig:KZ-fluxes}.
Three exponents $ x $ correspond to steady solutions of HIBE: the particle equipartition $ x_{eq}=0 $, the KZ particle cascade $ x_{\eta} $ and the KZ energy cascade $ x_{\epsilon} $.
As shown in Appendix \ref{A:KZ}, we know that $ \epsilon=0 $ on the particle cascade, $ \eta=0 $ on the energy cascade, whereas in the equipartition both fluxes are zero, i.e. $ \epsilon=\eta=0 $.
We also know that for large negative $ x $ (large positive slope) both fluxes must be negative, as such a steep unsteady spectrum would evolve to become less steep, toward equipartition.
Note that always $ x_{\eta} < x_{\epsilon} $, when $ x > 0 $.
Now we can sketch the particle and energy fluxes as function of the exponent $ x $
%and the fact that a very sharp initial distribution (negative slope for $ \omega $ less than the peak) will always try to reach the equipartition, the behavior can be qualitatively represented
as it is done in Fig. \ref{fig:KZ-fluxes}.
From this sketch, it can be easily understood that whenever  the condition $ x_{\eta} < x_{\epsilon} $ is valid, the particle flux will be positive and the energy flux will be negative, contradicting the Fj{\o}rtoft argument.
This means that one cannot match these formal KZ solutions, obtained for an infinite inertial range, to any physical forcing or dissipation at the ends of a large (but finite) inertial range.

What is then happening when fluxes have wrong direction? It has been observed in optical wave turbulence \cite{dyachenko:1992hc} that the pure KZ spectra are not established in these cases and one has to expect a mixed solution where both a flux and a thermal components are present.
Such mixed states are quite common for turbulent systems of different kinds, including strong Navier-Stokes turbulence  and have been named {\it warm cascades} \cite{nazarenko2004warm}.
Such cascades were obtained within the Leith model (which belongs to the class of the differential approximation models) as exact analytical solutions.

\section{Differential approximation model \label{DAM}}

Numerical integration of the Boltzmann collision integral is very challenging because the number of degrees of freedom grows as a polynomial. A great simplification comes from the isotopic assumption, which reduces the degrees of freedom from $ N^8 $ to $ N^2 $ ($ N $ is the number of points needed to describe the distribution). However spanning a large number of momentum scales is still difficult. For those reasons, some approximations to the kinetic equations were proposed in order to increase the range of modeled scales, see for example \cite{connaughton2009ns}.
% \red{[DAVIDE: maybe we can cite also Monte Carlo techniques?]}.

A great simplification is to replace the collision integral operator of the kinetic equation by a nonlinear differential operator which mimics the basic scalings of the original one and yields the same steady solutions. The HIBE then results in a nonlinear partial differential equation called the differential approximation model (DAM). Such models have been proposed to simulate turbulence in different research fields: for example in water waves \cite{hasselmann1985c2}, in nonlinear optics \cite{dyachenko:1992hc}, in strong Navier-Stokes turbulence \cite{leith1967,lvov2006di}, in Kelvin quantum turbulence \cite{boffetta2009mo},  in astrophysics (Kompaneets equation) \cite{peacock1999}, in semiconductors \cite{lvov1998quantum}. 
Replacing the integral operator by a differential one amounts to assuming locality of the scale interactions, which means the relevant distributions must be local for DAM to have a good predictive power. 
We mentioned in Section \ref{HIBE} that for hard sphere Boltzmann equation the pure KZ spectra are non-local and so no DAM would be advisable. However, we observed in some examples (Fig. \ref{fig:exampleEC} and Fig. \ref{fig:examplePC}) that the relevant solutions in this case are not pure KZ spectra but distributions which are close to MB, warm cascades, which appear to be local. Thus, we use the DAM for describing this system, after which we will validate our results by computing the full HIBE.

For the dual cascade systems, such as gravity water waves  \cite{zakharov1999diffusion}, nonlinear Schr{\"o}dinger equation \cite{dyachenko:1992hc}, two-dimensional hydrodynamic turbulence \cite{lvov2006di}, Kelvin waves  \cite{boffetta2009mo} or HIBE considered here, DAM has always the form of a dual conservation law,
\begin{equation}
\partial_t N(\omega, t)=\partial_{\omega\omega}R\left[n(\omega, t)\right],
\label{eq:DAM}
\end{equation}
where  $ R $ is a nonlinear second-order differential term whose details depend on the particular model.
This equation can be written as a continuity equation for the particle invariant, $$\partial_t N(\omega, t) + \partial_\omega\eta(\omega, t)= 0, $$ with the particle flux
\begin{equation}
\eta(\omega, t)=-\partial_\omega R\left[n(\omega, t)\right].
\label{eq:DAM-eta}
\end{equation}
Moreover, equation (\ref{eq:DAM}) can be written as a continuity equation for the energy \cite{dyachenko:1992hc}, 
$$ \partial_t \left[N(\omega, t)\omega\right] + \partial_\omega \epsilon(\omega, t) = 0, $$ 
with the energy flux
\begin{equation}
\epsilon(\omega, t)=R\left[n(\omega, t)\right]-\omega\partial_\omega R\left[n(\omega, t)\right].
\label{eq:DAM-epsilon}
\end{equation}

We are now able to find the functional $ R $ by requiring it to yield the MB distribution (\ref{eq:MB}) and the KZ spectra (\ref{eq:KZ}) as steady state solutions of DAM (\ref{eq:DAM}). These constraints lead to
\begin{equation}
R\left[n(\omega, t)\right]=-S \, \omega^{\frac{13}{2}} \, n^2(\omega, t) \, \partial_{\omega\omega} \log n(\omega, t),
\label{eq:R}
\end{equation}
where $ S $ is a constant. A formal derivation starting from the kinetic equation can be obtained following \cite{dyachenko:1992hc, lvov1998quantum}.
It is trivial to verify by substitution that KZ solutions (\ref{eq:KZ}) correspond to constant fluxes through scales. Namely, the KZ particle cascade has a constant particle flux and zero energy flux while the KZ energy cascade viceversa.
Let us again consider the the flux directions on the KZ distributions, but now using DAM.
Substituting power-law spectra $ n=c\, \omega^{-x} $ into (\ref{eq:R}), equations (\ref{eq:DAM-eta}) and (\ref{eq:DAM-epsilon}) yield
\begin{equation}
\begin{split}
& \eta = c^2 S \, x(9/2 -2x)\,  \omega^{7/2 -2x} \\
%\label{zeta-DA-BE} \\
%\label{epsilon-DA-BE}
& \epsilon = c^2 S \, x(11/2 -2x)\,  \omega^{9/2 -2x}.
\end{split}
\end{equation}
By plotting $ \eta $ and $ \epsilon $ as functions of the exponent $ x $ at fixed $ \omega $, we arrive again at Fig. \ref{fig:KZ-fluxes}.
Note that it is by using DAM such plot was obtained.
Once again we note that the particle and the energy fluxes on the respective KZ solutions ($ x=7/4 $ and $ x=9/4 $)
have wrong directions with respect to the Fj{\o}rtoft argument.

The beauty of the DAMs is the possibility to solve numerically the system  for wide frequency ranges and, therefore, to find clear scalings. In particular, such models are very efficient for finding constant steady flux solutions because they become  simple ordinary differential equations (ODEs). In the following we will present some analytical results for such steady states.

\subsection{Constant energy flux: direct cascade}

We will now find an ODE that describes a constant direct energy cascade $ \epsilon $ with no flux of particles, which we call ODE-$ \epsilon $. According to
Fj{\o}rtoft argument, this implies a large direct-cascade inertial range.
Putting  $ \eta=0 $ in (\ref{eq:DAM-eta}) and (\ref{eq:DAM-epsilon}), we have
\begin{equation}
\mbox{\underline{constant energy flux}} \ \ \ \Longrightarrow \ \ \  \epsilon=R(\omega, t)=const.
\label{eq:DAM-ce}
\end{equation}
%By considering a steady solution $ n(\omega) $, this means solving
Using (\ref{eq:R}), we arrive at the following Cauchy problem
\begin{equation}
\left\{
\begin{array}{l}
\epsilon = -S \, \omega^{\frac{13}{2}} n^2(\omega) \partial_{\omega\omega} \log n(\omega), \\
n(\omega_0)=n_0, \\
\partial_\omega n(\omega_0)=n'_0,
\end{array}
\right.
\label{eq:cauchy-dir}
\end{equation}
where we have chosen the boundary conditions fixing the values of the distribution and its derivative at the
same point $\omega_0$ (e.g. at the forcing scale) for ease of numerical solution.
%More physical conditions would be fixing $n=0$ at the dissipation scale (see the following for
%the behavior near the dissipation scale), but such a problem would be harder to compute
%because the ODE-$ \epsilon $ is highly nonlinear and non-autonomous.
%By analogy with the HIBE and problems due to direction of the flux we expect to find a warm cascade behavior.

If we solve numerically in $ \omega $-forward the ODE-$ \epsilon $ for different values of the energy flux we find curves presented in Fig. \ref{fig:energy_flux}. Here we do not want to discuss the details (it will be done widely in Section \ref{results}), but just remark that the solutions follow the MB distribution and suddenly change behavior going very fast to a zero value of the distribution. We will call this rapid change a {\it front solution}.

\subsubsection{Compact front behavior}

It is  possible to find a front solution for the equation (\ref{eq:DAM-ce}) describing the behavior near
the dissipation scale. Let us seek for a front solution
% propagating in the positive direction, a function that
which in the vicinity of a certain point $ \omega_{\max} $  behaves like $ n(\omega)=B\, (\omega_{\max}-\omega)^\sigma $.
If we plug this expression into (\ref{eq:DAM-ce}) and take the limit $ \omega \rightarrow \omega_{\max} $
we  find that to satisfy this equation in the leading order in $  (\omega_{\max} - \omega)$
we must have
%that, in order to have no singularities and to respect the positive flux condition, $ \epsilon > 0 $, it holds
\begin{equation}
\left\{
\begin{array}{l}
\sigma = 1\\
B=\sqrt{\frac{\epsilon}{S \, \omega_{\max}^{13/2}}}
\end{array}
\right. \ \ \ \Longrightarrow \ \ \ n(\omega)=\sqrt{\frac{\epsilon}{S \, \omega_{\max}^{13/2}}}\, (\omega_{\max}-\omega).
\label{eq:DAM-Be}
\end{equation}
Thus, the front solution is linear in the vicinity of $ \omega_{\max} $ with a  slope depending on the dissipation scale $ \omega_{\max} $ and the value of the energy flux $ \epsilon $. Note that the compact front behavior at the dissipation scale is typical for DAM.
%In the full integral kinetic equation this would be replaced with a fast super-polynomial (e.g. exponential) decay.
%We believe that the physical meaning of $ \omega_{\max} $ is to indicate at which scale the warm cascade ends and a dissipative term starts to act.
We will soon discover that $ \omega_{\max} $ is a very useful physical parameter which allows us to find a link between the temperature, the chemical potential and the energy flux in the forced-dissipated system.

\subsubsection{Kats-Kontorovich correction}

Lets summarize our preliminary observations. We expect a warm cascade, that is a distribution which contains both the flux and the thermal components.
We have also found that the solution has a compact front which arrests the cascade at the dissipation
 scale $ \omega_{\max} $.
% that depends on the flux and probably occurs in a region where the thermodynamic distribution become negligible.
We will now assume (verifying it later) that in the most of the inertial range the warm cascade solution is close to the thermodynamic MB distribution and  the  correction due to finite flux is small.
We then perform a qualitative matching of the flux-corrected MB distribution to the compact front, and thereby obtain a relation between $ \omega_{\max} $, $ T $ and $ A $ in (\ref{eq:MB}).
%two different behavior of the solution and we will like to match them.
To find the warm cascade solution in the inertial range,  we consider the Kats-Kontorovich (KK) correction
%{\color{red} [ref...?]} \blue{(cf. the Chapman-Enskog expansion technique)} \red{[DAVIDE: Pietro I don't think that KK corrections are similar to CE expansion..]} to the MB
to the Maxwell-Boltzmann 
distribution:
\begin{equation}
n(\omega)=n_{MB}(1+\tilde{n})=(1+\tilde{n})\, A\, e^{-\frac{\omega}{T}},
\label{eq:DAM-KK}
\end{equation}
where $ \tilde{n} $ is small, $ \tilde{n} \ll 1 $. By plugging this solution into (\ref{eq:DAM-ce}) and linearizing in $ \tilde{n} $ we end up with the following ODE-$ \epsilon $ for the correction
\begin{equation}
\epsilon\, \omega^{-\frac{13}{2}}A^{-2}\, e^{\frac{2\omega}{T}} = -S \, \partial_{\omega \omega}\tilde{n}.
\label{eq:KK-ef}
\end{equation}

\subsubsection{Matching}

We will now match the KK correction to
 the front solution. The basic idea is to force the KK solution to satisfy the $ n(\omega_{\max})=0 $ and to have at $ \omega_{\max} $ the same slope as the front solution.
%Then in the limit $ T \ll \omega_{\max} $ where the correction goes to zero.
Detailed calculation is  presented in Appendix \ref{ap:matchmax}. The prediction results in:
\begin{equation}
\epsilon=S\, \omega_{\max}^{\frac{9}{2}}\, A^2\, e^{-\frac{2\omega_{\max}}{T}}.
\label{eq:direct-pred}
\end{equation}
This relation is very important because it
gives an analytical relation between the thermodynamic quantities $ T $ and $ A $ in terms of the energy flux $ \epsilon $ and the dissipation scale  $ \omega_{\max} $. However we note that our matching is only qualitative, because the KK correction is supposed to be small which is not the case near the front.
%and to do the match with the front solution we have done the hypothesis that $ \tilde{n}=-1 $.
Thus, the relation (\ref{eq:direct-pred}) is approximate and we do not expect it to hold precisely.

%\subsubsection{Alternative approach to find $ \omega_{\max} $}
\subsubsection{Alternative approach to find \texorpdfstring{$ \omega_{\max} $}{omega max} }

Another simple way to find a prediction for the value of $ \omega_{\max} $ is the following. As we expect to observe a warm cascade, we can ask what will be the range where the thermal component will dominate the dynamics.
%A na{\"i}ve answer to that question can be obtained b
%By considering the initial conditions to satisfy the  thermodynamic solution $ n_{MB}(\omega_0)=A e^{-\frac{\omega_0}{T}} $,
We can simply assume that in most of the inertial range we will have a distribution $ n \simeq n_{MB} $.
%close to MB with the same $T$ and $A$.
Note that the MB distribution  always has a positive concavity,  $ \partial_{\omega\omega}n \ge 0 $.
On the other hand, we note that  our ODE-$ \epsilon $ can be re-written as
\begin{equation}
\label{eq:ode-second-der}
\partial_{\omega\omega}n=\frac{1}{n}\left[(\partial_\omega n)^2 - \frac{\epsilon}{S \, \omega^{\frac{13}{2}}} \right],
\end{equation}
from which it is clear that $ \partial_{\omega\omega}n  $ may change sign.
The point at which  $ \partial_{\omega\omega}n  =0$ can be considered as s boundary separating the MB range (with negligible flux correction) and the front solution (with large flux correction).
This boundary can be estimated by a simple substitution of the MB distribution to the r.h.s. of
(\ref{eq:ode-second-der}), which gives
%and that the Maxwell-Boltzmann  As $ n \ge 0 $ because it express a probability, in the hypothesis $ \partial_\omega n \simeq \partial_\omega n_{MB} $ we find
\begin{equation}
\epsilon = \frac{A^2\, S\, \omega^{\frac{13}{2}}\,e^{-\frac{2\omega}{T}}}{T^2}=g_{\epsilon}(\omega, A, T).
\label{eq:DAM-ge}
\end{equation}
%Considering that the front is very sharp, this boundary will be very close to $ \omega_{\max} $.
As this  relation contains the exponential factor which decays very fast (for $ \omega_{\max} \gg T$, see Appendix \ref{ap:matchmax}), it is natural to think that the range at which $ \epsilon $ becomes important appears very sharply and is very near to the point $ \omega_{\max} $.
Thus we arrive at the following  estimate,
\begin{equation}
\epsilon=\frac{A^2\, S\, \omega_{\max}^{\frac{13}{2}}\,e^{-\frac{2\omega_{\max}}{T}}}{T^2}.
\label{eq:DAM-gee}
\end{equation}

\subsection{Constant particle flux: inverse cascade}

In analogy of what has been done for the direct cascade, we now look for predictions in the inverse particle cascade $ \eta $ with no flux of energy.
The ODE-$ \eta $ that describes such a cascade is simple to obtain: by integrating equation (\ref{eq:DAM-eta}) once and putting $ \epsilon=0 $ in (\ref{eq:DAM-epsilon}), we have:
\begin{equation}
\mbox{\underline{constant particles flux}} \ \ \ \Longrightarrow \ \ \  \eta=-\frac{R(\omega, t)}{\omega}=const.
\label{eq:DAM-cp}
\end{equation}
%By considering a steady solution $ n(\omega) $, this means solving now
This yields the following  Cauchy problem,
\begin{equation}
\left\{
\begin{array}{l}
\eta = S \, \omega^{\frac{11}{2}} n^2(\omega) \partial_{\omega\omega} \log n(\omega), \\
n(\omega_0)=n_0, \\
\partial_\omega n(\omega_0)=n'_0.
\end{array}
\right.
\label{eq:cauchy-inv}
\end{equation}
This problem  is most naturally  solved backwards in the $ \omega $-space, as we are interested in the inverse cascade.
We seek for a solution having a particle flux going from high to low frequencies, i.e. $ \eta < 0 $ and for convenience we will make the substitution $ \eta \rightarrow -|\eta| $ in equation (\ref{eq:cauchy-inv}).
The Cauchy problem (\ref{eq:cauchy-inv}) is very similar to (\ref{eq:cauchy-dir}) with the only difference in the $ \omega $-scaling.
Thus we will use the same approach for studying it.

\subsubsection{Compact front behavior}

Let us find a front solution for the equation (\ref{eq:DAM-cp}). We now expect the front to be on the left edge of the (inverse cascade) inertial range, i.e. in the vicinity of a certain point $ \omega_{\min} <\omega_f $.
By plugging $ n(\omega)=B\, (\omega-\omega_{\min})^\sigma $ expression into (\ref{eq:DAM-cp})
and taking the limit $ \omega \rightarrow \omega_{\min} $,
%in order to have no singularities and respect the flux condition $ \eta > 0 $, it must be
in the leading order in $ (\omega - \omega_{\min} )$ we have
\begin{equation}
\left\{
\begin{array}{l}
\sigma = 1\\
B=\sqrt{\frac{|\eta|}{S \, \omega_{\min}^{11/2}}}
\end{array}
\right. \ \ \ \Longrightarrow \ \ \ n(\omega)=\sqrt{\frac{|\eta|}{S \, \omega_{\min}^{11/2}}}\, (\omega-\omega_{\min}).
\label{eq:DAM-Bp}
\end{equation}
Thus, the front solution for the inverse particle cascade is also linear in the vicinity of $ \omega_{\min} $, with a slope depending on  $ \omega_{\min} $ and the value of the particle flux $ \eta $.

\subsubsection{Kats-Kontorovich correction}

As previously supposed for the direct energy cascade, we expect in the most of the inverse-cascade range a
corrected thermodynamic spectrum and a front solution behavior at the left end of this range.
Let us evaluate the Kats-Kontorovich correction (\ref{eq:DAM-KK}), and after that match it to the front solution.
By plugging the expression (\ref{eq:DAM-KK}) into (\ref{eq:DAM-cp}) and linearizing in $ \tilde{n} $ we obtain the following ODE-$ \eta $ for the correction,
\begin{equation}
|\eta|\, \omega^{-\frac{11}{2}}\, A^{-2}\,  e^{\frac{2\omega}{T}} = -S \, \partial_{\omega \omega}\tilde{n}.
\label{eq:KK-pf}
\end{equation}

\subsubsection{Matching}
Again, we want to match the KK correction to the front solution.
The idea is very similar to the previously used for the direct cascade, except for the fact that now the limit
taken is $ \omega_{\min} \ll T $; for details refer to Appendix \ref{ap:matchmin}.
This results with the following  condition on the flux,
\begin{equation}
|\eta|=S \, \left(\frac{9}{2}\right)^2 A^2\, \omega_{\min}^{\frac{7}{2}}  .
\label{eq:inverse-pred}
\end{equation}

%\subsubsection{Alternative estimate of $ \omega_{\min} $}
\subsubsection{Alternative estimate of \texorpdfstring{$ \omega_{\min} $}{omega min} }

Again, we can obtain an alternative estimate for predicting the range of the warm cascade.
Let us rewrite the ODE-$ \eta $ as
\begin{equation}
\partial_{\omega\omega}n=\frac{1}{n}\left[(\partial_\omega n)^2 - \frac{|\eta|}{S \, \omega^{\frac{11}{2}}} \right].
\end{equation}
Keeping in mind that the MB
%Maxwell-Boltzmann 
distribution is always characterized by a positive concavity, i.e. $ \partial_{\omega\omega}n \ge 0 $, and considering the hypothesis $ \partial_\omega n \simeq \partial_\omega n_{MB} $ we find
\begin{equation}
|\eta| = \frac{A^2 \, S \, \omega^{\frac{11}{2}} \, e^{-\frac{2\omega}{T}}}{T^2}=g_{\eta}(\omega, A, T).
\label{eq:DAM-gp}
\end{equation}
%{\color{red} deriving from the Maxwellian is slowly varying $ \omega $ ?}
Similarly to what we have done for the inverse cascade, we now can suggest that the
change of concavity occurs near $ \omega_{\min} $. This results in
\begin{equation}
|\eta| = \frac{A^2 \, S \, \omega_{\min}^{\frac{11}{2}} \, e^{-\frac{2\omega_{\min}}{T}}}{T^2}.
\label{eq:DAM-gpp}
\end{equation}
However, we do not expect a good prediction as before because in this case the exponential term is not a rapidly varying function near $ \omega_{\min} $.

\subsection{Double cascade}

We have now all tools to study the double cascade process. Let us force at $ \omega_f $,  dissipate at $ \omega_{\max} $ and $ \omega_{\min} $, and consider the case $ \omega_{\min} \ll \omega_f \ll \omega_{\max} $.
If the forcing range is narrow, the simple relation $ \epsilon=\eta\, \omega_f $ holds for the fluxes.
Using this relation, and  combining (\ref{eq:direct-pred}) and (\ref{eq:inverse-pred}), we can estimate $ T $ and $ A $ in the system:
%\begin{equation}
%\left\{
\begin{equation}
\begin{split}
& T=\frac{2\omega_{\max}}{\frac{7}{2}\ln\frac{\omega_{\max}}{\omega_{\min}} +\ln\frac{\omega_{\max}}{\omega_f}-2\ln\frac{9}{2}},
\label{eq:T-mu} \\
& A=\frac{2}{9}\sqrt{\frac{|\eta|}{S \, \omega_{\min}^{7/2}}},
\end{split}
\end{equation}
%\right.
%\end{equation}
and, therefore, the chemical potential
\begin{equation}
\mu=T\left(\frac 1 {2} \ln\frac{S \, \omega_{\min}^{7/2}}{|\eta|}+\ln\frac{9}{2} \right).
\end{equation}
Note that the temperature appears to be independent of the fluxes and is completely controlled by the forcing and the dissipation scales. This means that increasing the forcing strength without moving $ \omega_f $ simply  adds more particles into the system with the energy per particle remaining the same.

\section{Numerical results  \label{results}}
In this Section we present the numerical results obtained by using the DAM and by integrating, at lower resolution, the HIBE. 
Our aim is to compare results for the warm cascade solutions of DAM, which has been devised as a local approximation of the integral collision operator, with direct numerical simulation of the full integro-differential equation (\ref{eq:fdHIBE}).

\subsection{DAM resutls}

We will first present some numerical experiments on integration of the Cauchy problems (\ref{eq:cauchy-dir}) and (\ref{eq:cauchy-inv}) in which we take for simplicity $ S=1 $.
Note that all numerical simulations can be performed without any loss of generality starting with a particular value $ \omega_0 $ because of re-scaling properties described in Appendix \ref{ap:rescaling}.

\subsubsection{Constant direct energy cascade}

\begin{figure}
\includegraphics[scale=1.6]{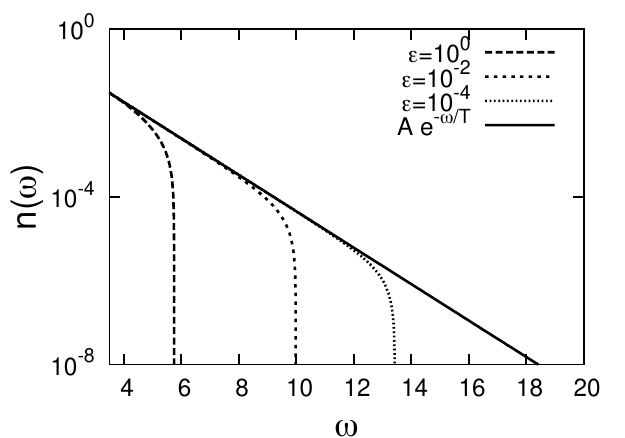}
\caption{DAM simulations of (\ref{eq:cauchy-dir}) for different constant energy flux $ \epsilon $ starting with the same initial condition at $ \omega_0=3.5 $ given by the MB distribution $ n_{MB}(\omega)=A e^{-\frac{\omega}{T}} $ where $ A=1 $ and $ T=1 $ (continuos line).
\label{fig:energy_flux}}
\end{figure}

In Fig. \ref{fig:energy_flux} we show the results obtained by integrating equation (\ref{eq:cauchy-dir}) with $ \omega_0=3.5 $ for different constant energy fluxes $ \epsilon $.
%\begin{figure}
%\includegraphics[scale=1.6]{DAM-estEC.pdf}
%\caption{DAM simulations of (\ref{eq:cauchy-dir}) for different constant energy flux $ \epsilon $ starting with the same initial condition at $ \omega_0=3.5 $ given by the MB distribution $ n_{MB}(\omega)=A e^{-\frac{\omega}{T}} $ where $ A=1 $ and $ T=1 $ (continuos line). Inset: plot in lin-log scale of the function $ g_\epsilon(\omega, A, T) $, see equation (\ref{eq:DAM-ge}), which qualitatively defines the thermodynamic regime of the solution.
%\label{fig:energy_flux}}
%\end{figure}
\begin{figure}
\includegraphics[scale=1.6]{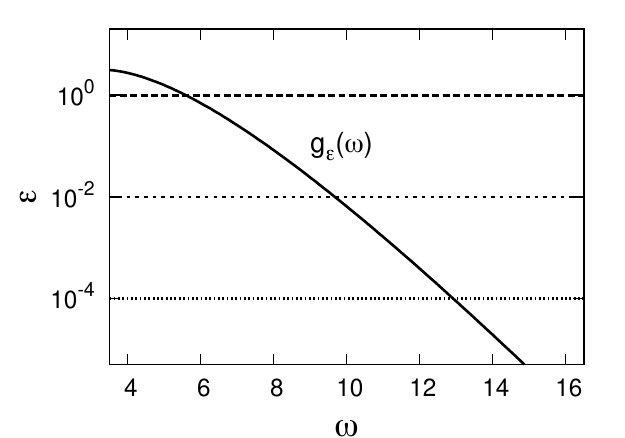}
\caption{Plot in lin-log scale of the function $ g_\epsilon(\omega, 1, 1) $, see equation (\ref{eq:DAM-ge}), which qualitatively defines the thermodynamic regime of the solution.
\label{fig:energy_flux-est}}
\end{figure}
As initial conditions, we choose the values of the spectrum $ n_0 $ and its slope $ n_0' $ from the MB distribution having $ A=1 $ and $ T=1 $. The solutions follow the thermodynamic solution (shown as a continuous line) until they rapidly deviate and reach the front in the vicinity of particular values of $ \omega_{\max} $.
This numerical experiment exhibits two important facts always observed in simulations performed with different initial conditions: the presence of a long transient in which the flux correction is negligible with respect to the thermodynamic MB distribution and the presence of a particular value $ \omega_{\max} $ at which $ n(\omega) $ goes to zero.
A lin-log plot of the function $ g_\epsilon(\omega, 1, 1) $, see equation (\ref{eq:DAM-ge}), is shown
%in the inset: 
in Fig. \ref{fig:energy_flux-est}:
intersection of this curve with horizontal lines at $ \epsilon=1 $, $ \epsilon=10^{-2} $ and $ \epsilon=10^{-4} $ marks the predicted cut-off frequencies for the respective flux values.
Agreement with the behavior in Fig. \ref{fig:energy_flux} is evident: the values of $ \omega_{\max} $ obtained with equation (\ref{eq:DAM-ge}) and Fig. \ref{fig:energy_flux-est} coincide with the observed values in Fig. \ref{fig:energy_flux} within 5\%.
Note that the peak of $ g_\epsilon(\omega, 1, 1,) $ is around $ \omega=3.5 $: this is why we set this value as initial condition $ \omega_0 $.

\begin{figure}
\includegraphics[scale=1.6]{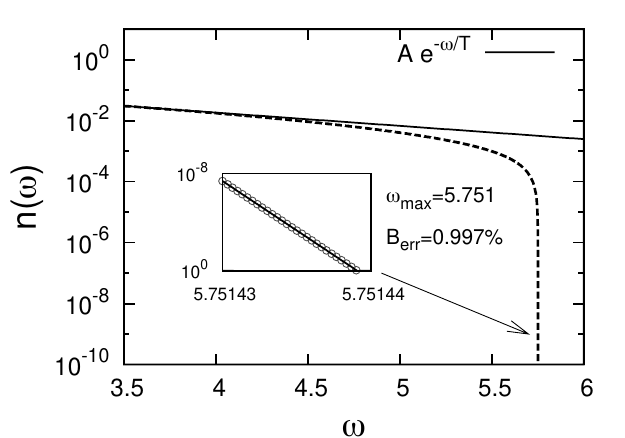}
\caption{ DAM simulation of ODE-$ \epsilon $ (\ref{eq:cauchy-dir}) with constant energy flux $ \epsilon=1 $ (dashed line).
The initial conditions in $ \omega_0=3.5 $ are set by the MB distribution with $ T=1 $ and $ A=1 $ (continuos line). The inset shows a zoom of numerical $ n(\omega) $ (dots) in the vicinity of the point $ \omega_{\max} $ where a linear fit is shown by continuos line.
\label{fig:omega_max}}
\end{figure}

In Fig. \ref{fig:omega_max} we present the results for a particular case with flux $ \epsilon=1 $.
We can appreciate the presence of warm cascade and the front solution near $ \omega_{\max} $. The linear behavior of the front is evident in the zoom near $ \omega_{\max} $ showed in the inset. Numerically we are able to measure $ \omega_{\max} $ and so evaluate $ B $ from equation (\ref{eq:DAM-Be}). The theoretical prediction agrees with the measured slope with the error $ B_{err} =0.997\%$.
The error is evaluated as $ B_{err}=|B_{meas}-B_{est}|/B_{meas} $ where $ B_{meas} $ is the measured linear coefficient and $ B_{est} $ is the one taken form relation (\ref{eq:DAM-Be}). 
In all other simulations performed with different values of $ \epsilon $ or different initial conditions, $ B_{err} $ is always within 5\%.

\begin{figure}
\includegraphics[scale=1.6]{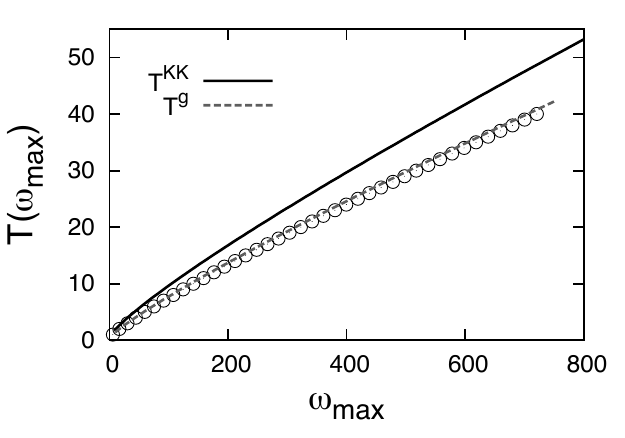}
\caption{Checking of predictions in DAM constant energy flux cascade (\ref{eq:cauchy-dir}): the points represent the temperature of the initial condition $ T $ with respect to the measured $ \omega_{\max} $. Solid line is the matching relation
%(\ref{eq:ener-T-w})
(\ref{eq:direct-pred})
while dashed one is obtained from (\ref{eq:DAM-gee}). \label{fig:prediction}}
\end{figure}

We now check numerically the validity of the matching prediction (\ref{eq:direct-pred}) by taking different initial condition $ n_{MB}(\omega_0=3.5) $ varying $ T $ and keeping $ A=1 $ and $ \epsilon=1$: results are plotted in Fig. \ref{fig:prediction}.
It is evident from the figure that the predicted temperature (continuous black line) evaluated from relation (\ref{eq:direct-pred})
%\begin{equation}
%T^{KK}=\frac{2\omega_{\max}}{\frac{9}{2}\log\omega_{\max}+2\log A-\log \eta}=2\frac{\omega_{\max}+\mu}{\frac{9}{2}\log\omega_{\max}-\log \eta}
%\label{eq:ener-T-w}
%\end{equation}
is an overestimation of the numerical results (dots) and the error is around 10\%.
%\red{[DAVIDE: We can observe that, if the chemical potential is negligible with respect to $ \omega_{\max} $, the direct energy cascade is completely responsable to the temperature of the system. SN: Davide, explain -I do not understand what you said.]}
Finally prediction for the alternative temperature relation (\ref{eq:DAM-gee}) is plotted with gray dashed line: it appears to give a better estimation
than relation (\ref{eq:direct-pred}).

\subsubsection{Constant particle cascade}

\begin{figure}
\includegraphics[scale=1.6]{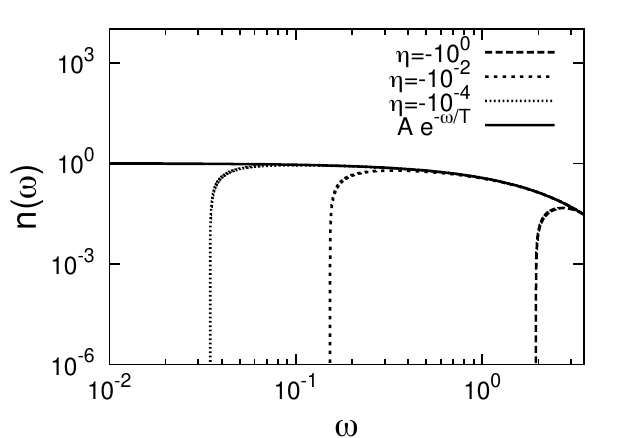}
\caption{DAM simulations of (\ref{eq:cauchy-inv}) for different constant inverse particle flux $ \eta $ starting with the same initial condition at $ \omega_0=3.5 $ given by the Boltzmann distribution $ n_{MB}(\omega)=A e^{-\frac{\omega}{T}} $ where $ A=1 $ and $ T=1 $ (plotted with continuos line).
\label{fig:DAM-predp}}
\end{figure}

\begin{figure}
\includegraphics[scale=1.6]{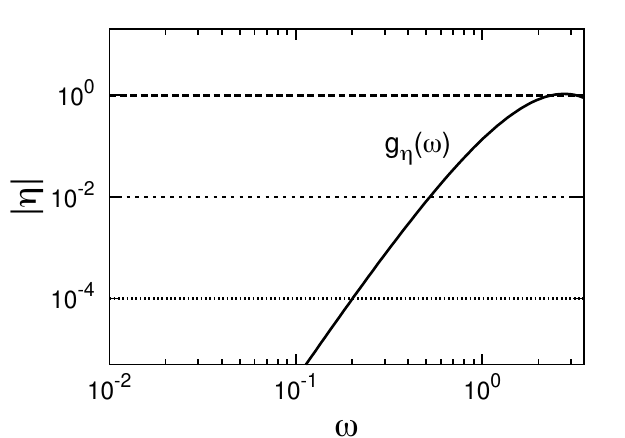}
\caption{Plot of the function $ g_\eta(\omega, 1, 1) $, see equation (\ref{eq:DAM-gp}).
\label{fig:DAM-predp-est}}
\end{figure}

We now investigate the  inverse particle cascade by solving  Cauchy problem (\ref{eq:cauchy-inv}) going $ \omega $-backward. In Fig. \ref{fig:DAM-predp} we show numerical results obtained by taking initial conditions at $ \omega_0 $ from MB distribution $ n_{MB}(\omega)=A e^{-\frac{\omega}{T}} $ with $ T=1 $, $ A=1 $. As in the case of constant energy flux, here the warm cascade range is wider for smaller  flux
values. 
We also observe fronts in vicinities of cutoff points $ \omega_{\min} $.
In Fig. \ref{fig:DAM-predp-est} we show the function $ g_\eta(\omega, 1, 1) $ which represents the prediction of the thermodynamic range (\ref{eq:DAM-gp}).
Qualitative front values of results in Fig. \ref{fig:DAM-predp} show poor agreement with this na{\"i}ve estimation. 
%\begin{figure}
%\includegraphics[scale=1.6]{DAM-estPC.pdf}
%\caption{DAM simulations of (\ref{eq:cauchy-inv}) for different constant inverse particles flux $ \eta $ starting with the same initial condition at $ \omega_0=3.5 $ given by the Boltzmann distribution $ n_{MB}(\omega)=A e^{-\frac{\omega}{T}} $ where $ A=1 $ and $ T=1 $ (plotted with continuos line). Inset: plot of the function $ g_\eta(\omega, A, T) $, see equation (\ref{eq:DAM-gp}).
%\label{fig:DAM-predp}}
%\end{figure}

\begin{figure}
\includegraphics[scale=1.6]{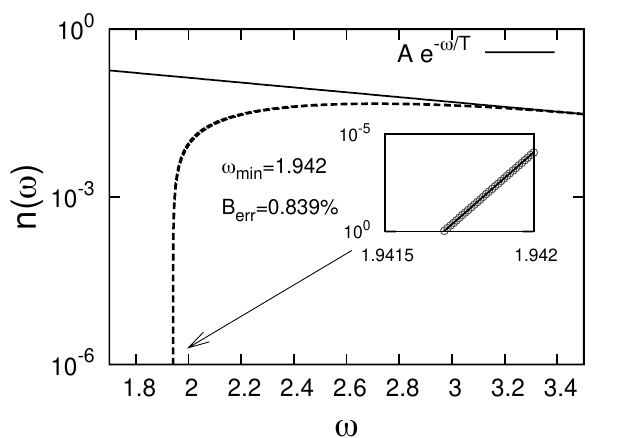}
\caption{DAM simulation of (\ref{eq:cauchy-inv}) with $ \eta=-1 $ (dashed line) starting with initial condition at $ \omega_0=3.5 $ given by the MB distribution $ n_{MB}(\omega)=A e^{-\frac{\omega}{T}} $ where $ A=1 $ and $ T=1 $ (plotted with continuos line). Inset: lin-lin scale zoom in the vicinity $ \omega_{\min} $  (dots) where the best linear fit is presented with a continuous line.
\label{fig:DAM-omega_min}}
\end{figure}

The front solution is analysed in detail in Fig. \ref{fig:DAM-omega_min} where we choose  the particular case with $ \eta=-1 $. The linear behavior is demonstrated in the inset. Moreover a numerical estimation of $ \omega_{\min} $ lets us evaluate $ B $, see equation (\ref{eq:DAM-Bp}). The error $ B_{err} $ is presented in the figure; for all other  simulations we have performed   $ B_{err} $ remained within 4\%.

\begin{figure}
\includegraphics[scale=1.6]{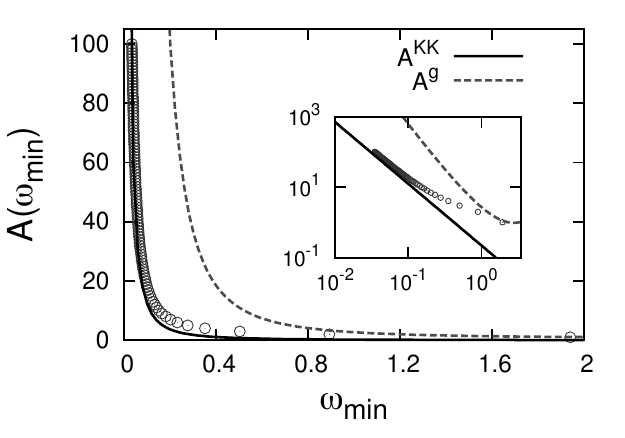}
\caption{Checking predictions in DAM constant inverse flux cascade (\ref{eq:cauchy-inv}): the points represent the thermodynamic quantity $ A $ with respect to the measured $ \omega_{\min} $. Continuos line is the KK matching prediction given in equation (\ref{eq:inverse-pred}) while dashed one is obtained from (\ref{eq:DAM-gpp}).
\label{fig:ener-A-w}}
\end{figure}

Finally we check KK matching prediction for the thermodynamic quantity $ A $ with respect to $ \omega_{\min}  $ presented in equation (\ref{eq:inverse-pred}): results are showed in Fig. \ref{fig:ener-A-w}.
In this case the $ KK $ analytical prediction (continuous line) underestimates the numerical data while the estimation (\ref{eq:DAM-gpp}) is completely out of range (dashed line). However the scaling $ A \sim \omega_{\min}^{-7/4} $ of KK prediction tends to be reached for small values of $ \omega_{\min} $, where $ \omega_{\min} \ll T $.

\subsubsection{Double cascade}

\begin{figure}
\includegraphics[scale=1.6]{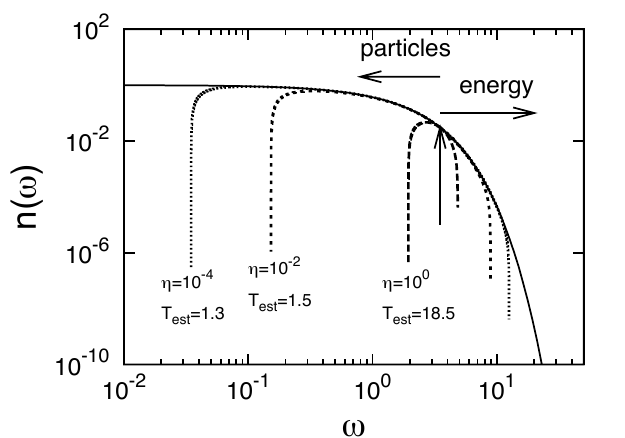}
\caption{DAM double cascade simulations of equations (\ref{eq:cauchy-dir}) and (\ref{eq:cauchy-inv}) for three different values of the particle flux $ \eta $ (and consequently of the energy flux $ \epsilon=\eta\,  \omega_f $). The initial condition are  taken at $ \omega_f=3.5 $ from the MB distribution with $ T=1 $ and $ A=1 $. Measuring $ \omega_{\min} $ and $ \omega_{\max} $ in each case we estimate of the temperature $ T_{est} $ from prediction (\ref{eq:T-mu}).
\label{fig:double}}
\end{figure}

An example of double cascade is presented in Fig. \ref{fig:double} where we set the forcing at $ \omega_f=\omega_0=3.5 $.
We show here three cases where the particle fluxes are respectively $ \eta=-1 $, $ \eta=-10^{-2} $ and $ \eta=-10^{-4} $. Measuring $ \omega_{\min} $ and $ \omega_{\max} $ for each case we are able to estimate the temperature $ T_{est} $ from prediction (\ref{eq:T-mu}).
Results do not agree with the expected temperature (the initial conditions set it at $ T=1 $) but they approach this value for bigger ranges, i.e. when the condition $  \omega_{\min} \ll \omega_f \ll \omega_{\max} $ is better satisfied (see for example the case $ \eta=-10^{-4} $).

\subsection{HIBE results \label{hibeCode}}

We now to present results of the direct simulation of HIBE with the full Boltzmann collision integral and compare them with predictions obtained by DAM. 
As we have mentioned above, the evaluation of (\ref{eq:HIBE}) is numerically challenging and it is nowadays practically impossible  to simulate such wide $ \omega $-space ranges  as we have done using the DAM.
In the present work, we will always use a low resolution of 101 points by considering $ \omega \in \left[0, \omega_{cutoff}\right] $ and taking a uniform distribution with $ \Delta\omega=\frac{\omega_{cutoff}}{100} $.
We have checked that the numerical solutions are mesh independent by taking a finer mesh, 201 points, and comparing the solution of one critical case.
%\red{[PIETRO: essentially there is no check that the numerical solutions are mesh independent. A sensitivity analysis is mandatory (at least for one critical case). Essentially a second (sligtly) finer mesh must be considered for comparing the numerical results, in order to check how much they depend on the considered mesh]} 

The $ \delta $-function in (\ref{eq:HIBE}) defines a resonant manifold over which the integrand need to be evaluated; numerically it is a set of discrete resonant conditions $ \mathcal{M}=\left\{\omega_1, \omega_2, \omega_3, \omega_4\right\} \ \ / \ \  \omega_1+\omega_2 = \omega_3+\omega_4 $ which can be pre-computed.
Note that the dissipation at high wave numbers is chosen to satisfy $ \omega_{\max} \le \omega_{cutoff}/2 $ in order to prevent ultraviolet bottleneck effects. The time evolution is performed by using the Euler scheme.
Further details on numerical methods for solving the HIBE and a simple code can be found in \cite{Asinari20101776}.

\subsubsection{Direct cascade study}

We first analyze the direct energy cascade by putting the forcing scale near  the low-$ \omega $ dissipation scale in order to have a wider direct inertial range. Numerical results for these final steady states were previously presented as examples in Fig. \ref{fig:exampleEC} and Fig. \ref{fig:ke-flux}.
%\begin{figure}
%\includegraphics[scale=1.6]{spectra.pdf}
%\caption{Steady states of HIBE in linear-logarithmic scale obtained by choosing $ \omega_{cutoff}=200$, $ \omega_{\min}=5 $, $ \omega_f=21 $ and $ \omega_{\max}=95. $ for different values of the forcing. Inset: total energy $ E(t) $ in the system as a function of time.\label{fig:ke-flux}}
%\end{figure}
We concentrate now only on the last one: here we kept fixed $ \omega_{\min}=5 $, $ \omega_f=21 $ and $ \omega_{\max}=95 $ and varied the forcing coefficient, i.e. the fluxes $ \eta $ and $ \epsilon $.
We were claiming that the temperature of the systems is the same because qualitatively the distributions have identical slopes. 
Moreover we observed in all the examples that left and right branch chemical potentials and temperatures can be defined by the forcing scale.
%In particular we can notice that measured temperature is always greater in the left zone while $ A $ is bigger to the right. 

\begin{figure}
\includegraphics[scale=1.6]{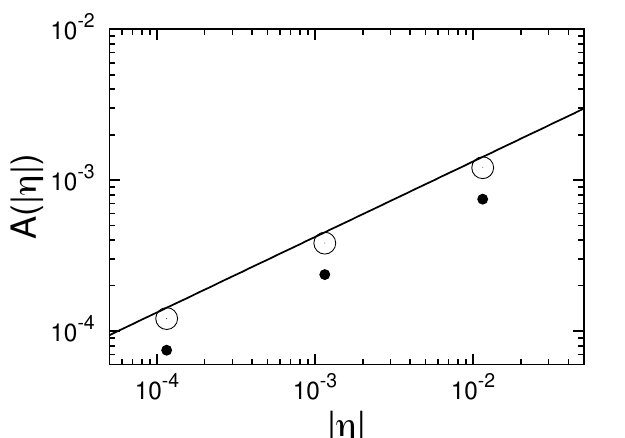}
\caption{Results for the fitted values of the thermodynamic quantity $ A=e^{-\frac{\mu}{T}} $ plotted against the forcing levels, as obtained in the  simulations shown in Fig. \ref{fig:ke-flux}:
 % in logarithmic-logarithmic plot with respect to different value of the particle flux $ \eta $
the values of   $ A_L $ are shown by filled circles while $ A_R $ - by empty ones, the blue line refers to prediction (\ref{eq:T-mu}) with $ S=1 $. \label{fig:ke-T-N}}
\end{figure}

With these previous DAM results in mind we have measured $ A $ and $ T $ in three examples presented in Fig. \ref{fig:ke-flux}: the results are shown in Fig. \ref{fig:ke-T-N} and are compared to analytical predictions (\ref{eq:T-mu}).
%\begin{figure}
%\includegraphics[scale=1.6]{A-T_flux.pdf}
%\caption{Results for the thermodynamic quantities plotted against the forcing levels, as obtained in the  simulations shown in Fig. \ref{fig:ke-flux}. The main plot shows the fitted values of $ A=e^{-\frac{\mu}{T}} $:
 % in logarithmic-logarithmic plot with respect to different value of the particle flux $ \eta $
%the values of   $ A_L $ are shown by filled circles while $ A_R $ - by empty ones, the blue line refers to prediction (\ref{eq:T-mu}) with $ S=1 $. The inset shows temperatures $ T_L $ (filled triangles) and $ T_R $ (empty ones);
% for different energy flux $ \epsilon $;
%the dashed line is the analytical prediction. \label{fig:ke-T-N}}
%\end{figure}
\begin{figure}
\includegraphics[scale=1.6]{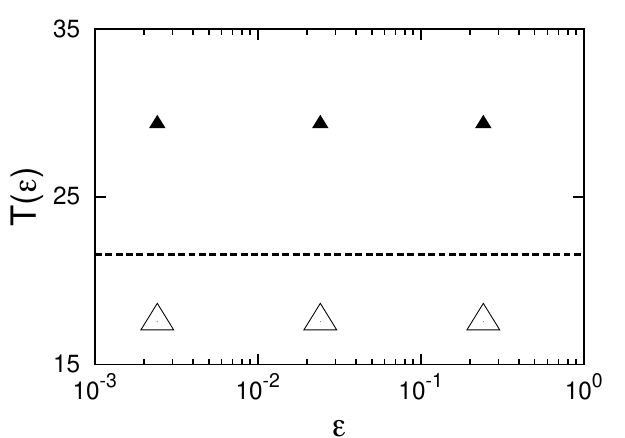}
\caption{Measured temperatures $ T_L $ (filled triangles) and $ T_R $ (empty ones) obtained in the simulations shown in Fig. \ref{fig:ke-flux}.
The dashed line is the analytical prediction (\ref{eq:T-mu}). \label{fig:ke-T}}
\end{figure}
As expected the quantity $ A \sim \sqrt{\eta} $ but the line (in log-log plot) is shifted with respect to the interval between $ A_L $ and $ A_R $, represented respectively with filled and empty circles. 
However, the theoretical prediction is much closer to  $ A_R $, which is natural because the right inertial interval is wider than the left one. 
In fact, the agreement of $ A_R $ with the theory is quite good considering  the presence of the undefined constant $ S $ in the theoretical prediction. 
The temperature is shown in Fig. \ref{fig:ke-T}: even though $ T_L $ and $ T_R $ are different they both appear to be forcing independent, as predicted. The temperature evaluated from relation (\ref{eq:T-mu}): temperature (dashed line) stands in between of these values, and closer to  $ T_R $, which, again, is natural because the right inertial interval is wider.

\begin{figure}
\includegraphics[scale=1.6]{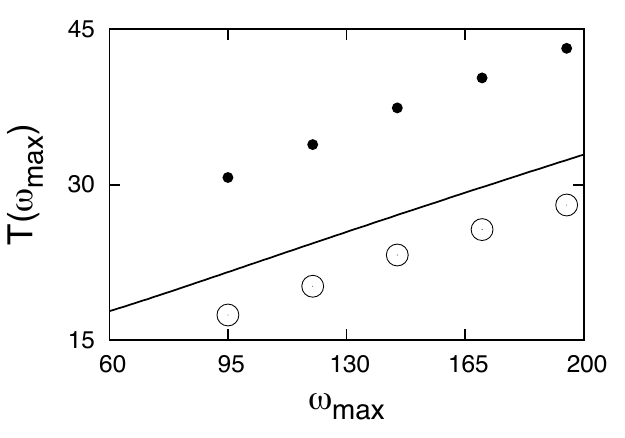}
\caption{Temperature in different steady state keeping the forcing constant and varying the  dissipation scale $ \omega_{\max} $: big empty circles correspond to the temperature $ T_R $ on the right of the forcing scale, whereas  small filled circles to the left side, $ T_L $. The continuous line is the prediction (\ref{eq:T-mu}). \label{fig:T-omax}}
\end{figure}

We have also analyzed sensitivity of the temperature to varying the high-$ \omega $ dissipation range and results are presented in Fig. \ref{fig:T-omax}.
Keeping the forcing constant  and changing the value of $ \omega_{max} $ the system reaches steady states characterized by different temperatures $ T_L $ (filled circles) and $ T_R $ (empty circles). The prediction (\ref{eq:T-mu}), shown by the continuous line, is in between of the two temperatures and is closer to  $ T_R $ - again due to the wider right range.

\subsubsection{Inverse cascade study}

\begin{figure}
\includegraphics[scale=1.6]{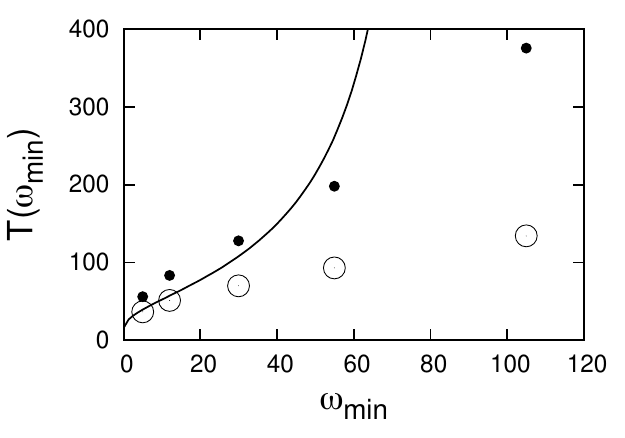}
\caption{Temperature in different steady state keeping the forcing constant and varying the dissipation scale $ \omega_{\min} $: big empty circles correspond to the temperature $ T_R $ on the right of the forcing scale, whereas  small filled circles to the left side, $ T_L $. The continuous line is the prediction (\ref{eq:T-mu}). \label{fig:T-omin}}
\end{figure}

\begin{figure}
\includegraphics[scale=1.6]{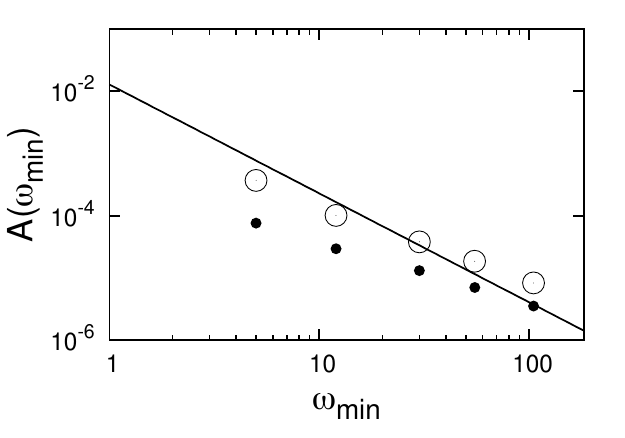}
\caption{Thermodynamic amplitude $ A $ for different $ \omega_{\min} $ and fixed forcing in log-log scales in numerical simulations of HIBE. Filled circles corresponds to $ A_L $ while empty circles - $ A_R $. 
The continuos line is formula (\ref{eq:T-mu}) with $ S=1 $. \label{fig:N-omin}}
\end{figure}

Finally, we have performed some simulations putting the forcing scale near the dissipation at high $ \omega $'s in order to study the inverse cascade process. In this case too, as reported in Fig. \ref{fig:examplePC}, we observe two different values of thermodynamic quantities on the left and on the right from the forcing.
Here we are able to study the scaling of the thermodynamic quantities $ T $ and $ A $ with respect to changes of the small-$ \omega $  dissipation scale $ \omega_{\min} $. Results for $ T $ are shown in Fig. \ref{fig:T-omin}
and for $ A $ in Fig. \ref{fig:N-omin}, with the ``left" quantities shown by filled circles and the ``right" ones by empty circles. There is a reasonably good agreement of $ T $ with the prediction (\ref{eq:T-mu})
for small $ \omega_{\min} $. This is natural because smaller $ \omega_{\min} $ corresponds to larger  
inverse cascade inertial range and also because the prediction is valid when $ \omega_{\min} \ll T $. 

On the other hand, for $A$ the prediction (\ref{eq:T-mu}) is in better agreement with 
the data at large $ \omega_{\min} $ with $ \omega_{\min} \sim \omega_f $.
%\begin{figure}
%\includegraphics[scale=1.6]{A_omega_min.pdf}
%\caption{Thermodynamic amplitude $ A $ for different $ \omega_{\min} $ and fixed forcing in log-log scales in numerical simulations of HIBE. Filled circles corresponds to $ A_L $ while empty circles - $ A_R $. 
%The continuos line is formula (\ref{eq:T-mu}) with $ S=1 $. The inset shows in log-linear scale the ratio of the measured inverse particle flux over the imposed one $ \eta_L/\eta $ (empty triangles) with respect to $ \omega_{\min} $; the continuous line correspond to the imposed flux while the dashed one follows the finite range correction (\ref{eq:etaL}). \label{fig:N-omin}}
%\end{figure}
\begin{figure}
\includegraphics[scale=1.6]{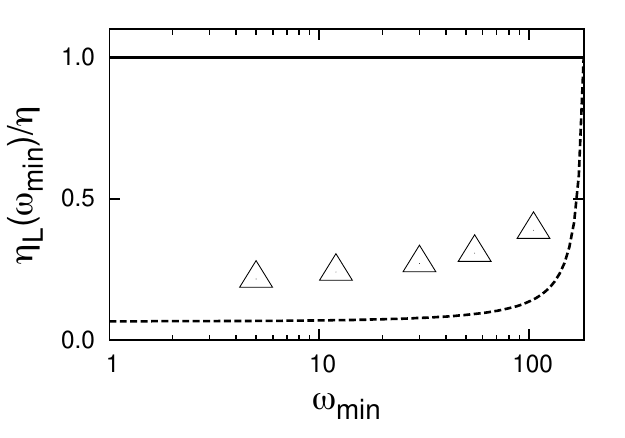}
\caption{The plot shows in log-linear scale the ratio of the measured inverse particle flux over the total one $ \eta_L/\eta $ (empty triangles) with respect to $ \omega_{\min} $; the continuous line correspond to the total flux while the dashed one follows the finite range correction (\ref{eq:etaL}). \label{fig:eta-omin}}
\end{figure}
This is due to two possible reasons.
First, we underline that the agreement can be made more suitable since the analytical prediction contains  the undefined order-one parameter $ S $ which could be adjusted to better fit the numerical results. 
Second, the particle flux which defines $ A $ in relation (\ref{eq:T-mu}) can be smaller due to finite range effects. Indeed, following \cite{Newell2001520}, the ratio of the leftward particle flux to the total particle production rate is estimated in
\begin{equation}
\eta_L = \eta \frac{\omega_{\max}-\omega_f}{\omega_{\max}-\omega_{\min}}.
\label{eq:etaL}
\end{equation}
This equation, in addition to other corresponding to rightward fluxes in the cited paper, states that for the particle flux to be mostly to the left inertial ranges in both directions must be large (note that this is also the condition of validity of the Fj{\o}rtoft argument).
Fig. \ref{fig:eta-omin} shows the behavior of the normalised left measured flux $ \eta_L/\eta $ (empty triangles) with respect to $ \omega_{\min} $.
We can clearly see that the measured particle flux is indeed much smaller than the one imposed by the forcing term (continuos line), around one third of it. This is in quite good agreement with the finite range prediction (\ref{eq:etaL}) plotted with dashed line.
Similar reasoning can be made for corrections on $ A(|\eta|) $ in the case of direct cascade example in Fig. \ref{fig:ke-T-N}.

%\red{ SN: Do you have a figure for the temperature for the inverse cascade? For variable forcing?} 

\section{Conclusions \label{conclusions}}
In the present paper we investigated stationary turbulent states in the isotropic Boltzmann kinetic equation for hard spheres. This was done by looking for steady nonequilibrium states in open systems, that is when forcing and dissipation mechanisms are present. Analogies with similar results of wave turbulence theory suggest the manifestation of a {\it warm cascade}, i.e. a constant direct flux of energy and inverse flux of particles on background of thermodynamic Maxwell-Boltzmann distribution. This is a consequence of wrong flux directions in KZ solutions with respect to the Fj{\o}rtoft argument.

We have built an ad-hoc differential approximation model to easily simulate the cascade processes. Indeed, this simplification allowed us to reach a wide range of scales inaccessible by solving the isotropic Boltzmann kinetic equation directly. Simulations show the presence of a warm cascade with approximately the MB shape followed by a sharp front for both energy and particle cascades. We have physically interpreted $ \omega_{\min} $ and $ \omega_{\max} $ as intrinsic dissipation scales at low and high $ \omega $'s which are necessary to establish the steady state. Moreover, we have found analytical predictions relating the particle and energy fluxes, forcing and dissipations scales to the thermodynamic quantities of the system. In particular we have shown that the temperature is independent of the amplitude of the fluxes but only depends on the forcing and dissipation scales.

We have then compared the theoretical predictions and the numerical results obtained with the differential approximation model with simulations of the complete isotropic Boltzmann kinetic equation. Even though the resolution for the latter was limited by the available computational power, the results are comparable and in good agreement with the analytical predictions. In particular we have verified that the steady state is characterized by a warm cascade where a fitted thermodynamic Maxwell-Boltzmann distribution has been used to measure temperature and chemical potential of the system. We observe, in agreement with our analytical predictions, that the temperature is completely defined by the forcing and dissipation scales and does not depend on the fluxes.

%\red{[DAVIDE: Importance in what field? Physical examples of the system, granular fluids?]}

We hope that this work may open some perspectives towards understanding nonequilibrium steady states and their net currents (fluxes) by cross-fertilization with the weak turbulence theory.

\begin{acknowledgments}
We would like to thank Guido Boffetta, Colm Connaughton, Filippo De Lillo, Stefano Musacchio, Al Osborne, and Arturo Viero for fruitful discussions. Simulations were performed on computational resources founded by the Office of Naval Research (ONR). Finally, we are grateful to the Gnu Scientific Library (GSL) developers for providing free software which has been used for simulations.  
\end{acknowledgments}


\appendix

\section{Three-dimensional  \texorpdfstring{$ \delta $}{delta}-function angular average \label{delta3d}}

The angular average of the four-wave linear momentum conservation $ \delta(\mathbf{k}_{12}^{34}) = \delta(\mathbf{k}_1+\mathbf{k}_2-\mathbf{k}_3-\mathbf{k}_4) $ is evaluated by splitting it into two $ \delta $-functions of three particle collision. This results in
\begin{eqnarray}
\int_{\Omega} \delta(\mathbf{k}_{12}^{34}) d\Omega_{1234} & = & \int_{\Omega} \int_{\mathbf{k}_{\min}}^{\mathbf{k}_{\max}} \delta(\mathbf{k}_1+\mathbf{k}_2-\mathbf{k}) \delta(\mathbf{k}_3+\mathbf{k}_4-\mathbf{k}) d\mathbf{k} d\Omega_{1234} \nonumber \\
& = & \int_{k_{\min}}^{k_{\max}}  \left[  \int_{\Omega}  \delta(\mathbf{k}_1+\mathbf{k}_2-\mathbf{k})d\Omega_{12} \right] \left[  \int_{\Omega} \delta(\mathbf{k}_3+\mathbf{k}_4-\mathbf{k})d\Omega_{34} \right] k^{d-1} dk d\Omega \nonumber \\
& = & 4\pi \int_{k_{\min}}^{k_{\max}} \frac{1}{2kk_1k_2} \frac{1}{2kk_3k_4} k^2 dk = \frac{2\pi}{k_1k_2k_3k_4}\min(k_1, k_2, k_3, k_4),
\end{eqnarray}
where geometrically $ k_{\min}=|k_1-k_2|=|k_3-k_4| $ and $ k_{\max}=k_1+k_2=k_3+k_4 $. For details about the integration of three particle $ \delta $-function see Appendices in \cite{zakharov41kst}.

\section{Kolmogorov-Zakharov solutions for general HIBE\label{A:KZ}}
The Boltzmann collision integral $ I_{coll} $ is defined as
\begin{eqnarray}
I_{coll}(\mathbf{x}, \mathbf{k}_1, t) & = & \int_{-\infty}^\infty \Gamma_{12}^{34} \left[n(\mathbf{x}, \mathbf{k}_3, t) n(\mathbf{x}, \mathbf{k}_4, t) - n(\mathbf{x}, \mathbf{k}_1, t) n(\mathbf{x}, \mathbf{k}_2, t)\right] \nonumber \\ 
& & \times \delta(\mathbf{k}_1+\mathbf{k}_2-\mathbf{k}_3-\mathbf{k}_1) \delta(|\mathbf{k}_1|^2+|\mathbf{k}_2|^2-|\mathbf{k}_3|^2-|\mathbf{k}_4|^2) d\mathbf{k}_{234},
\end{eqnarray}
where the two $ \delta $-functions assure the conservation of the linear momentum and kinetic energy.
In the isotropic case it is convenient to move in the energy domain $ \omega_i=|\mathbf{k}_i|^2 \in [0, +\infty) $ and so the HIBE results in
\begin{equation}
I(\omega_1)=\int_0^\infty S_{12}^{34} (n_3n_4 - n_1n_2) \delta(\omega_{12}^{34}) d\omega_{234},
\label{int_234}
\end{equation}
where $ I(\omega_1) = \Omega_1 \, I_{coll}(\mathbf{x}, \omega_1, t) \omega_1^{\frac{d-1}{2}} \left| \frac{dk_1}{d\omega_1} \right|  $ and we use for brevity $ n_i = n(\omega_i)=n(\mathbf{x}, |\mathbf{k}_i|^2, t) $, and $ \delta(\omega_{12}^{34})=\delta(\omega_1+\omega_2-\omega_3-\omega_4) $. 
The functional $ S $ is
\begin{equation}
S_{12}^{34} = \frac{1}{16} (\omega_1\omega_2\omega_3\omega_4)^{\frac{d}{2}-1} \langle \Gamma_{12}^{34} \delta(\mathbf{k}_1+\mathbf{k}_2-\mathbf{k}_3-\mathbf{k}_4)\rangle_{\Omega}
\end{equation}
and the operator $ \langle \cdot \rangle_{\Omega} $ states for the integration over solid angles.
%\begin{eqnarray}
%S_{12}^{34} & = & 4\pi \Omega_0 (\omega_1\omega_2\omega_3\omega_4)^{\frac{d-1}{2}} \left|\frac{dk_1}{d\omega_1}\right|\left|\frac{dk_2}{d\omega_2}\right| \left|\frac{dk_3}{d\omega_3}\right| \left|\frac{dk_4}{d\omega_4}\right| \langle \Gamma_{12}^{34} \delta(\mathbf{k}_1+\mathbf{k}_2-\mathbf{k}_3-\mathbf{k}_4)\rangle \nonumber \\
%& = & \frac{\pi\Omega_0}{4} (\omega_1\omega_2\omega_3\omega_4)^{\frac{d}{2}-1} \langle \Gamma_{12}^{34} \delta(\mathbf{k}_1+\mathbf{k}_2-\mathbf{k}_3-\mathbf{k}_4)\rangle.
%\end{eqnarray}
%The functional $ S $ satisfies all the discrete symmetries of the collisional kernel $ \Gamma_{12}^{34} $.
It is important for the following to estimate the homogeneity degree of $ S $. Supposing that the collisional kernel scales as $ \Gamma_{\lambda(12)}^{\lambda(34)}=\lambda^{2\beta} \Gamma_{12}^{34} $, we have
\begin{equation}
S_{\lambda(12)}^{\lambda(34)}=\lambda^{4\left(\frac{d}{2}-1\right)+2\beta-\frac{d}{2}}S_{12}^{34}=\lambda^{\frac{3d}{2}+2\beta-4}S_{12}^{34}.
\end{equation}
Moreover its behavior at the boundaries of integration is
\begin{equation}
\begin{split}
& \lim_{\omega_i \rightarrow +\infty} S_{12}^{34} \sim \omega_i^{d-2+\tau_1} \\
& \lim_{\omega_i \rightarrow 0^+} S_{12}^{34} \sim \omega_i^{\frac{d}{2}-1+\tau_2}
\end{split}
\end{equation}
if we assume that 
\begin{equation}
\lim_{\omega_i \rightarrow +\infty} \langle \Gamma_{12}^{34} \delta(\mathbf{k}_1+\mathbf{k}_2-\mathbf{k}_3-\mathbf{k}_4)\rangle_{\Omega}  \sim \omega_i^{\tau_1}
\end{equation}
and
\begin{equation}
\lim_{\omega_i \rightarrow 0^+} \langle \Gamma_{12}^{34} \delta(\mathbf{k}_1+\mathbf{k}_2-\mathbf{k}_3-\mathbf{k}_4)\rangle_{\Omega} \sim \omega_i^{\tau_2}
\end{equation}
(note that for $ \omega_i \rightarrow \infty $ also another $ \omega_j $ must go to infinity due to the $ \delta $-function).
%Using the ansatz
%\begin{equation}
%\left\{
%\begin{array}{l}
%\lim_{\omega_i \rightarrow +\infty} \langle \Gamma_{12}^{34} \delta(\mathbf{k}_1+\mathbf{k}_2-\mathbf{k}_3-\mathbf{k}_4)\rangle  \sim \omega_i^{\tau_1} \\
%\lim_{\omega_i \rightarrow 0^+} \langle \Gamma_{12}^{34} \delta(\mathbf{k}_1+\mathbf{k}_2-\mathbf{k}_3-\mathbf{k}_4)\rangle \sim \omega_i^{\tau_2}
%\end{array}
%\right.
%\end{equation}
%we find (note that for $ \omega_i \rightarrow \infty $ also another $ \omega_j $ must go to infinity due to the $ \delta $-function)
%\begin{equation}
%\left\{
%\begin{array}{l}
%\lim_{\omega_i \rightarrow +\infty} S_{12}^{34} \sim \omega_i^{d-2+\tau_1} \\
%\lim_{\omega_i \rightarrow 0^+} S_{12}^{34} \sim \omega_i^{\frac{d}{2}-1+\tau_2}
%\end{array}
%\right.
%\end{equation}

In the following we will suppose that the particle distribution function follows the power-law distribution $ n(\omega) = A \, \omega^{-\nu} $ and so
\begin{equation}
I(\omega_1) = A^2 \int_0^\infty S_{1(3+4-1)}^{34} \left[ \omega_3^{-\nu} \omega_4^{-\nu} - \omega_1^{-\nu}\left(\omega_3+\omega_4-\omega_1\right)^{-\nu} \right] \Theta(\omega_3+\omega_4-\omega_1) \, d\omega_{34},
\label{int_34}
\end{equation}
%\begin{eqnarray}
%I(\omega_1) & = & A^2 \int_0^\infty S_{12}^{34} (\omega_3^{-\nu} \omega_4^{-\nu} - \omega_1^{-\nu}\omega_2^{-\nu})\delta(\omega_{12}^{34}) d\omega_{234} \nonumber \\
%& = & A^2 \int_0^\infty S_{1(3+4-1)}^{34} \left[ \omega_3^{-\nu} \omega_4^{-\nu} - \omega_1^{-\nu}\left(\omega_3+\omega_4-\omega_1\right)^{-\nu} \right] \Theta(\omega_3+\omega_4-\omega_1) d\omega_{34},
%\label{int_34}
%\end{eqnarray}
where $ \Theta $ is the Heaviside step function.

\subsection{Kolmogorov-Zakharov solutions}
We will present the Kolmogorov-Zakharov solutions of the collision integral using the method presented by Balk in \cite{balk2000kzs}. The collision integral, without any loss of generality, can be rewritten as
\begin{equation}
I(\omega_1)= A^2 \omega_1^{-1-\mu} \int_0^\infty S_{12}^{34} \, (\omega_3^{-\nu} \omega_4^{-\nu} - \omega_1^{-\nu}\omega_2^{-\nu}) \, (\omega_1\omega_2\omega_3\omega_4) \, \omega_1^{\mu} \, \delta(\omega_{12}^{34}) \,\frac{d\omega_2}{\omega_2}\frac{d\omega_3}{\omega_3}\frac{d\omega_4}{\omega_4}
\end{equation}
where the exponent
\begin{equation}
\mu=2\nu+1-2\beta-\frac{3d}{2}
\end{equation}
is chosen in order to have zero as homogeneity coefficient of the integrand (excluding the differentials $ \frac{d\omega_i}{\omega_i} $).  If the integral converges, Balk proved that is possible to interchange the three integration index in the integrand with the fourth one, $ \omega_1 $. Thanks to the symmetric properties of the collision kernel we can write
\begin{eqnarray}
I(\omega_1) & = & \frac{A^2 \omega_1^{-1-\mu}}{4} \int_0^\infty S_{12}^{34} \, (\omega_3^{-\nu} \omega_4^{-\nu} - \omega_1^{-\nu}\omega_2^{-\nu}) \, (\omega_1\omega_2\omega_3\omega_4) \nonumber \\
& & \times (\omega_1^{\mu}+\omega_2^{\mu}-\omega_3^{\mu}-\omega_4^{\mu}) \, \delta(\omega_{12}^{34}) \, \frac{d\omega_2}{\omega_2}\frac{d\omega_3}{\omega_3}\frac{d\omega_4}{\omega_4},
\label{int_balk}
\end{eqnarray}
which clearly vanishes for $ \mu=0 $ or $ \mu=1 $. This corresponds to the condition on the exponent
\begin{equation}
\begin{split}
& \nu_0=\nu|_{\mu=0}=\frac{3d-2}{4}+\beta \\
& \nu_1=\nu|_{\mu=1}=\frac{3d}{4}+\beta.
\end{split}
\end{equation}
Note that first KZ solution for HIBE were presented in \cite{kats1975}.

\subsection{Convergence of the integral (locality condition)}
The locality of interactions is guaranteed by the convergence of the collision integral. We then investigate the possible values of $ \nu $ which assure the convergence around the integrand singularities.

%\subsubsection{Limit $ \omega_3\rightarrow\infty $}
\subsubsection{Limit \texorpdfstring{$ \omega_3\rightarrow \infty $}{omega3 tends to infinity}}

In the limit of $ \omega_3\rightarrow\infty $ we can approximate $ (\omega_3+\omega_4-\omega_1)^{-\nu} = \omega_3^{-\nu} - \nu\omega_3^{-\nu-1}(\omega_4-\omega_1) + O(\omega_3^{-\nu-2}) $
%\begin{equation}
%(\omega_3+\omega_4-\omega_1)^{-\nu} = \omega_3^{-\nu} - \nu\omega_3^{-\nu-1}(\omega_4-\omega_1) + O(\omega_3^{-\nu-2})
%\end{equation}
at the second order. The argument in the square brackets of (\ref{int_34}) results in
\begin{equation}
[...] \simeq \omega_3^{-\nu} \left[\omega_4^{-\nu} - \omega_1^{-\nu} + \nu \omega_1^{-\nu}\omega_3^{-1}(\omega_4-\omega_1)\right].
\end{equation}
%\begin{eqnarray}
%[...] & = & \omega_3^{-\nu}\omega_4^{-\nu} - \omega_1^{-\nu} \omega_3^{-\nu} + \nu \omega_1^{-\nu}\omega_3^{-\nu-1}(\omega_4-\omega_1) \nonumber \\
%& = & \omega_3^{-\nu} \left[\omega_4^{-\nu} - \omega_1^{-\nu} + \nu \omega_1^{-\nu}\omega_3^{-1}(\omega_4-\omega_1)\right]
%\end{eqnarray}
As a consequence, when $ \nu>0 $, the integrand for large $ \omega_3 $ goes like $ \frac{\omega_4^{-\nu}-\omega_1^{-\nu}}{\omega_3^{\nu-d+2-\tau1}} $ and so the convergence condition is
\begin{equation}
\nu>d-1+\tau_1 .
\end{equation}

%\subsubsection{Limit $ \omega_3\rightarrow 0^+ $}
\subsubsection{Limit \texorpdfstring{$ \omega_3\rightarrow 0^+ $}{omega3 tends to zero}}

In the limit of $ \omega_3\rightarrow 0^+ $ we can approximate $ (\omega_3+\omega_4-\omega_1)^{-\nu} = (\omega_4-\omega_1)^{-\nu} -\nu\omega_3(\omega_4-\omega_1)^{-\nu-1} + O(\omega_3^2) $
%\begin{equation}
%(\omega_3+\omega_4-\omega_1)^{-\nu} = (\omega_4-\omega_1)^{-\nu} -\nu\omega_3(\omega_4-\omega_1)^{-\nu-1} + O(\omega_3^2)
%\end{equation}
at the second order. The argument in the square brackets of (\ref{int_34}) results in
\begin{equation}
[...] = \omega_3^{-\nu} \left[\omega_4^{-\nu} - \omega_1^{-\nu}\omega_3^{\nu}(\omega_4-\omega_1)^{-\nu} + \nu \omega_1^{-\nu}\omega_3^{\nu+1}(\omega_4-\omega_1)^{-\nu-1}\right] .
\end{equation}
%\begin{eqnarray}
%[...] & = & \omega_3^{-\nu}\omega_4^{-\nu} - \omega_1^{-\nu} (\omega_4-\omega_1)^{-\nu} + \nu \omega_1^{-\nu}\omega_3(\omega_4-\omega_1)^{-\nu-1} \nonumber \\
%& = & \omega_3^{-\nu} \left[\omega_4^{-\nu} - \omega_1^{-\nu}\omega_3^{\nu}(\omega_4-\omega_1)^{-\nu} + \nu \omega_1^{-\nu}\omega_3^{\nu+1}(\omega_4-\omega_1)^{-\nu-1}\right]
%\end{eqnarray}
So, when $ \nu>0 $, the integrand for small $ \omega_3 $ goes like $ \frac{\omega_4^{-\nu}}{\omega_3^{\nu-\frac{d}{2}+1-\tau2}} $ and so the convergence condition is
\begin{equation}
\nu<\frac{d}{2}+\tau_2 .
\end{equation}
Analogue condition holds for the singularity $ (\omega_3+\omega_4-\omega_1)^{-\nu} \rightarrow 0^+ $.

\subsection{Constant fluxes}
The solutions $ n(\omega)=A \, \omega^{-\nu_0} $ and $ n(\omega)=A \, \omega^{-\nu_1} $ correspond, respectively, to constant flux of particle and energy. To demonstrate this fact we perform the substitution $ \omega_i=\omega_1\xi_i $ $ \forall \, i\neq1 $ in the equation (\ref{int_balk}) which results, recalling the homogeneity of the integrand function, in
\begin{eqnarray}
I(\omega_1) & = & \frac{A^2 \, \omega_1^{-1-\mu}}{4} \int_{\Delta} S_{1\xi_2}^{\xi_3\xi_4} \, (\xi_3^{-\nu} \xi_4^{-\nu} - \xi_2^{-\nu}) \, (\xi_2\xi_3\xi_4) \nonumber \\
& & \times (1+\xi_2^{\mu}-\xi_3^{\mu}-\xi_4^{\mu}) \, \delta(1+\xi_2-\xi_3-\xi_4) \, d\xi_{234}=\frac{A^2 \, \omega_1^{-1-\mu}}{4} \, U(\mu)
\label{int_dil}
\end{eqnarray}
The integral $ U(\mu) $ is now performed over the triangle $ \Delta $ in the $ \xi_3\times\xi_4 $ space satisfying the conditions $ 0 \le \xi_i \le 1 $ and $ \xi_4 \ge 1-\xi_3 $, without any dependence on $ \omega_1 $.
%, see Fig. \ref{triangle}
%\begin{figure}
%\includegraphics[scale=0.8]{triangle.pdf}
%\caption{Integration domain $ \Delta $ of integral $ U(\mu) $, see equation (\ref{int_dil}).\label{triangle}}
%\end{figure}

\subsubsection{Flux of particles}
The flux of particles is defined as
\begin{equation}
Q(\omega)=-\int_0^\omega I(\omega_1) \, d\omega_1 = -\frac{A^2 \, U(\mu)}{4} \int_0^\omega \omega_1^{-1-\mu} d\omega_1 = \frac{A^2 \, U(\mu) \, \omega^{-\mu}}{4 \mu}.
\end{equation}
If $ \mu=1 $ the flux is zero while in the case $ \mu=0 $ it is indeterminate. By applying the De l'H\^opital rule in the latter case we find
\begin{equation}
Q(\omega)\left.\right|_{\mu=0} = \frac{A^2}{4} \int_{\Delta} S_{1\xi_2}^{\xi_3\xi_4} \,  (\xi_3^{-\nu_0} \xi_4^{-\nu_0} - \xi_2^{-\nu_0}) \, (\xi_2\xi_3\xi_4) \, \ln{\left(\frac{\xi_2}{\xi_3 \xi_4}\right)} \, \delta(1+\xi_2-\xi_3-\xi_4) \, d\xi_{234}
\end{equation}
%\begin{eqnarray}
%Q(\omega)\left.\right|_{\mu=0} & = &-\frac{A^2}{4} \left[\frac{dU(\mu)}{d\mu}\omega_1^{-\mu} - \mu U(\mu)\omega_1^{-\mu-1}\right]_{\mu=0}=-\frac{A^2}{4} \left. \frac{dU(\mu)}{d\mu} \right|_{\mu=0}  \\
%& = & \frac{A^2}{4} \int_{\Delta} S_{1\xi_2}^{\xi_3\xi_4} (\xi_3^{-\nu_0} \xi_4^{-\nu_0} - \xi_2^{-\nu_0}) (\xi_2\xi_3\xi_4) \ln{\left(\frac{\xi_3\xi_4}{\xi_2}\right)} \delta(1+\xi_2-\xi_3-\xi_4) d\xi_{234} \nonumber
%\end{eqnarray}
The integrand, and so the sign of the particle flux, is always negative for $ \nu_0 > 0 $. This is clear by looking at the sign of every factors in the integral: all are trivially positive except $ (\xi_3^{-\nu_0} \xi_4^{-\nu_0} - \xi_2^{-\nu_0}) $ and $ \ln{\left(\frac{\xi_2}{\xi_3 \xi_4}\right)} $. Recalling that $ (1-\xi_3)(1-\xi_4) \ge 0 $ and $ \xi_2=\xi_3+\xi_4-1 $ we have
\begin{equation}
0 \le (1-\xi_3)(1-\xi_4) = \xi_3\xi_4 - \xi_3 -\xi_4 +1=\xi_3\xi_4 - \xi_2 \Longrightarrow \xi_3\xi_4 \ge \xi_2,
\end{equation}
which leads to $ \ln{\left(\frac{\xi_2}{\xi_3 \xi_4}\right)} \le 0 $ and $ (\xi_3^{-\nu_0} \xi_4^{-\nu_0} - \xi_2^{- \nu_0}) \le 0 $ (for positive $ \nu_0 $).
As a consequence $ Q(\omega) \ge 0 $, that is the particle flux goes from low to high frequencies. 

%We analyse the convergence of the flux in the neighborhood of $ \xi_2 \rightarrow 0^+ $. The integral can be approximated to
%\begin{equation}
%\lim_{\xi_2 \rightarrow 0^+} Q(\omega)\left.\right|_{\mu=0} \sim \int_0^\epsilon \xi_2^{\frac{d}{2}+\tau_2-\nu_0} \ln\xi_2 d\xi_2 = \left. \frac{\xi_2^{\frac{d}{2}+\tau_2-\nu_0+1}}{\frac{d}{2}+\tau_2-\nu_0+1}  \left(\ln\xi_2 - \frac{1}{\frac{d}{2}+\tau_2-\nu_0+1}\right) \right|_0^\epsilon
%\end{equation}
%which converges only if
%\begin{equation}
%\nu_0 < \frac{d}{2}+\tau2+1
%\end{equation}

\subsubsection{Flux of energy}
The flux of energy is
\begin{equation}
P(\omega)=-\int_0^\omega I(\omega_1) \, \omega_1 \, d\omega_1 = -\frac{A^2 \, U(\mu)}{4} \int_0^\omega \omega_1^{-\mu} d\omega_1 = -\frac{A^2 \, U(\mu) \, \omega_1^{1-\mu}}{4 (1-\mu)}
\end{equation}
and is null when $ \mu=0 $ while indeterminate in the case $ \mu=1 $. Again applying the De l'H\^opital rule we have
\begin{eqnarray}
P(\omega)\left.\right|_{\mu=1} & = & \frac{A^2}{4} \int_{\Delta} S_{1\xi_2}^{\xi_3\xi_4} \, (\xi_3^{-\nu_1} \xi_4^{-\nu_1} - \xi_2^{-\nu_1}) \, (\xi_2\xi_3\xi_4) \\
& & \times \left[\xi_2\ln(\xi_2) -\xi_3\ln(\xi_3)-\xi_4\ln(\xi_4) \right] \, \delta(1+\xi_2-\xi_3-\xi_4) \, d\xi_{234} \nonumber
\end{eqnarray}
%\begin{eqnarray}
%P(\omega)\left.\right|_{\mu=1} & = & \frac{A^2}{4} \left[\frac{dU(\mu)}{d\mu}\omega_1^{1-\mu} + (1-\mu) U(\mu)\omega_1^{-\mu}\right]_{\mu=1}= \frac{A^2}{4} \left. \frac{dU(\mu)}{d\mu} \right|_{\mu=1}  \\
%& = & \frac{A^2}{4} \int_{\Delta} S_{1\xi_2}^{\xi_3\xi_4} (\xi_3^{-\nu_1} \xi_4^{-\nu_1} - \xi_2^{-\nu_1}) (\xi_2\xi_3\xi_4) \nonumber \\
%& & \times \left[\xi_2\ln(\xi_2) -\xi_3\ln(\xi_3)-\xi_4\ln(\xi_4) \right] \delta(1+\xi_2-\xi_3-\xi_4) d\xi_{234} \nonumber
%\end{eqnarray}
As previously discussed, the term $ (\xi_3^{-\nu_1} \xi_4^{-\nu_1} - \xi_2^{-\nu_1}) \le 0 $ for every $ \nu_1>0 $.
Differently, the factor $ \left[\xi_2\ln(\xi_2) -\xi_3\ln(\xi_3)-\xi_4\ln(\xi_4) \right] $ is always positive but here the demonstration is not so trivial as in the previous case and for a complete discussion see \cite{kats1976}. So $ P(\omega) \le 0 $, which means that the energy flux goes from high to low frequencies.

%Again we estimate the convergence criteria of the flux in the neighborhood of $ \xi_2 \rightarrow 0^+ $. The integral can be approximated to
%\begin{equation}
%\lim_{\xi_2 \rightarrow 0^+} Q(\omega)\left.\right|_{\mu=1} \sim \int_0^\epsilon \xi_2^{\frac{d}{2}+\tau_2-\nu_1} d\xi_2 = \left. \frac{\xi_2^{\frac{d}{2}+\tau_2-\nu_1+1}}{\frac{d}{2}+\tau_2-\nu_1+1}  \right|_0^\epsilon
%\end{equation}
%which converges if
%\begin{equation}
%\nu_1 < \frac{d}{2}+\tau2+1
%\end{equation}

%\subsection{General conditions}
%Taking into account all conditions of locality and convergence of the fluxes we find that the interval of the parameter $ \beta $ for the two Kolmogorov-Zakharov solutions. Please note that $ \tau_1 $ and $ \tau_2 $ are still functions of $ \beta $. In the case of constant flux of particles we have $ \nu_0=\frac{3d-2}{4}+\beta $ and so
%\begin{equation}
%d-1+\tau_1<\nu_0<\frac{d}{2}+\tau_2
%\Longrightarrow
%\left\{
%\begin{array}{l}
%\beta>\frac{d-2+4\tau_1}{4} \\
%\beta<\frac{2-d+4\tau_2}{4}
%\end{array}
%\right.
%\label{cond_0}
%\end{equation}
%For the constant energy flux solution we have $ \nu_1=\frac{3d}{4}+\beta $ and
%\begin{equation}
%d-1+\tau_1<\nu_1<\frac{d}{2}+\tau_2
%\Longrightarrow
%\left\{
%\begin{array}{l}
%\beta>\frac{4\tau_1+d-4}{4} \\
%\beta<\frac{4\tau_2-d}{4}
%\end{array}
%\right.
%\label{cond_1}
%\end{equation}
%It is trivial to see that all these flux solutions can exist only if $ \tau_2 > \frac{d}{2}-1+\tau_1 $.

%\section{Matching Kats-Kontorovich to $ \omega_{\max} $ front solution\label{ap:matchmax}}
\section{Matching Kats-Kontorovich to \texorpdfstring{$ \omega_{\max} $}{omega max} front solution\label{ap:matchmax}}

We will here find the match between the KK correction and the front solution for the ODE-$ \epsilon $.
%for that case is
%\begin{equation}
%\epsilon \omega^{-\frac{13}{2}}A^{-2} e^{\frac{2\omega}{T}} = -\partial_{\omega \omega}\tilde{n}.
%\end{equation}
We make the hypothesis that the front occurs for $ \omega_{\max} \gg T $ and so it is reasonable to think that the term $ \omega^{-\frac{13}{2}} $ in equation (\ref{eq:KK-ef}) it is slowly varying with respect to $ e^{\frac{2\omega}{T}} $. So by integrating twice in $ \omega $ (\ref{eq:KK-ef}) and match to the front we get the Cauchy problem
\begin{equation}
\left\{
\begin{array}{l}
\epsilon \omega^{-\frac{13}{2}}\frac{T^2}{4}A^{-2}e^{\frac{2\omega}{T}}=-S \, \tilde{n}+c_1(\omega-\omega_{\max})+c_2 \\
\tilde{n}(\omega_{\max})=-1 \\
\partial_\omega \tilde{n}(\omega_{\max})=B A e^{\frac{\omega_{\max}}{T}}
\end{array}
\right.
\end{equation}
where the first condition assures that $ n(\omega_{\max})=0 $ and the second that the front behavior is linear with slope $ B=-\epsilon^{\frac{1}{2}}\omega_{\max}^{-\frac{13}{4}} $ found in equation (\ref{eq:DAM-Be}). The integration constants are then
\begin{equation}
\left\{
\begin{array}{l}
c_2=-S+\epsilon \, \omega_{\max}^{-\frac{13}{2}} \, \frac{T^2}{4} \, A^{-2} \,e^{\frac{2\omega_{\max}}{T}} \\
c_1=S^{-\frac{1}{2}} \, \epsilon^{\frac{1}{2}}\,\omega_{\max}^{-\frac{13}{4}}\,A^{-1}\,e^{\frac{\omega_{\max}}{T}} - \epsilon\,\omega_{\max}^{-\frac{13}{2}}\,\frac{T}{2}\, A^{-2}\, e^{\frac{2\omega_{\max}}{T}}.
\end{array}
\right.
\end{equation}
We will now match this solution in the regime where $ T \ll \omega_{\max} $ and the flux is negligible with respect to the thermodynamic solution. In this regime, by assuming the scaling relation $ \epsilon \sim \omega_{\max}^{\frac{9}{2}}A^{2}e^{-\frac{2\omega_{\max}}{T}} $, the coefficients results in
\begin{equation}
\left\{
\begin{array}{l}
c_2 \simeq -S \\
c_1 \simeq S^{-\frac{1}{2}}\,\epsilon^{\frac{1}{2}}\omega_{\max}^{-\frac{13}{4}} A^{-1} e^{\frac{\omega_{\max}}{T}}
\end{array}
\right.
\end{equation}
and so the smallness of the correction reads as
\begin{equation}
\tilde{n}(\omega)=0=-\epsilon \omega^{-\frac{13}{2}}\frac{T^2}{4} A^{-2} e^{\frac{2\omega}{T}} + (\omega-\omega_{\max})S^{-\frac{1}{2}} \epsilon^{\frac{1}{2}}\omega_{\max}^{-\frac{13}{4}} A^{-1} e^{\frac{\omega_{\max}}{T}}-S.
\end{equation}
Finally, considering that $ \omega \ll \omega_{\max} $, we recover and validate the relation
\begin{equation}
\epsilon=S \,\omega_{\max}^{\frac{9}{2}}\, A^2\, e^{-\frac{2\omega_{\max}}{T}}.
\end{equation}

%\section{Matching Kats-Kontorovich to $ \omega_{\min} $ \label{ap:matchmin} front solution}
\section{Matching Kats-Kontorovich to \texorpdfstring{$ \omega_{\min} $}{omega min} front solution \label{ap:matchmin}}

The KK correction for that case is given by equation (\ref{eq:KK-pf}).
%\begin{equation}
%|\eta| \omega^{-\frac{11}{2}}A^{-2} e^{\frac{2\omega}{T}} = -\partial_{\omega \omega}\tilde{n}.
%\end{equation}
We will consider the limit $ \omega \ll T $ and so $ e^{\frac{2\omega}{T}} \simeq 1 $. By integrating twice in $ \omega $ we get the Cauchy problem
\begin{equation}
\left\{
\begin{array}{l}
|\eta| \, \omega^{-\frac{7}{2}}\,\frac{4}{63}\, A^{-2}=-S\,\tilde{n}+c_1(\omega-\omega_{\min})+c_2 \\
\tilde{n}(\omega_{\min})=-1 \\
\partial_\omega \tilde{n}(\omega_{\min}) = |\eta|^{\frac{1}{2}}\,S\,\omega_{\min}^{-\frac{11}{4}}\,A^{-1}
\end{array}
\right.
\end{equation}
with the initial conditions chosen in order to match with the front solution. The integration constants result in
\begin{equation}
\left\{
\begin{array}{l}
c_2=-S +|\eta|\, \omega_{\min}^{-\frac{7}{2}}\,\frac{4}{63}\, A^{-2} \\
c_1=-S^{\frac{1}{2}}\,|\eta|^{\frac{1}{2}}\omega_{\min}^{-\frac{11}{4}} A^{-1} +|\eta|\omega_{\min}^{-\frac{9}{2}}\frac{2}{9} A^{-2}.
\end{array}
\right.
\end{equation}
We assume and guess that the particles flux scales as $ |\eta| \sim \omega_{\min}^{\frac{7}{2}} \, A^2 $. Now, in the regime $ \omega \gg \omega_{\min} $ were the correction is negligible we have
\begin{equation}
\tilde{n}(\omega)=0=-|\eta| \, \omega^{-\frac{7}{2}}\,\frac{4}{63}\, A^{-2} +  c_1(\omega-\omega_{\min}) + c_2 \simeq \omega \, c_1.
\end{equation}
So, finally, we impose that $ c_1=0 $ to get the condition on the flux
\begin{equation}
|\eta|=S\, \left(\frac{9}{2}\right)^2 A^2\, \omega_{\min}^{\frac{7}{2}}  .
\end{equation}

\section{Scaling properties of DAM\label{ap:rescaling}}

The constant energy flux DAM (\ref{eq:DAM-ce}) and the constant particles flux DAM (\ref{eq:DAM-cp}) can be generally written as
\begin{equation}
c = - S \omega^p n^2(\omega) \partial_{\omega\omega} \log n(\omega)
\end{equation}
where the exponent is respectively $ p=13/2 $ and $ p=11/2 $ and $ c $ is a constant that represent the considered flux.
Lets now analyze the rescaling properties of that equation by the following change of variables
\begin{equation}
\left\{
\begin{array}{l}
c=\lambda^\alpha \bar c \\
\omega=\lambda^\beta \bar \omega \\
n=\lambda^\gamma \bar n.
\end{array}
\right.
\end{equation}
After some easy algebra we find that the system is invariant if
\begin{equation}
\alpha=(p-2)\beta+2\gamma.
\end{equation}
As a consequence we can establish how the thermodynamic quantities defined by the Maxwell-Boltzmann distribution $ n_{MB}(\omega)=A e^{-\frac{\omega}{T}}=e^{-\frac{\omega+\mu}{T}} $ vary: the temperature $ T $ scales as $ \omega $ and so $ T=\lambda^\beta \bar T $, while the chemical potential $ \mu $ scales as $ \mu=\lambda^\beta \gamma \bar\mu $.

\bibliography{references}

\begin{thebibliography}{32}
\expandafter\ifx\csname natexlab\endcsname\relax\def\natexlab#1{#1}\fi
\expandafter\ifx\csname bibnamefont\endcsname\relax
  \def\bibnamefont#1{#1}\fi
\expandafter\ifx\csname bibfnamefont\endcsname\relax
  \def\bibfnamefont#1{#1}\fi
\expandafter\ifx\csname citenamefont\endcsname\relax
  \def\citenamefont#1{#1}\fi
\expandafter\ifx\csname url\endcsname\relax
  \def\url#1{\texttt{#1}}\fi
\expandafter\ifx\csname urlprefix\endcsname\relax\def\urlprefix{URL }\fi
\providecommand{\bibinfo}[2]{#2}
\providecommand{\eprint}[2][]{\url{#2}}

\bibitem[{\citenamefont{Bustamante et~al.}(2005)\citenamefont{Bustamante,
  Liphardt, and Ritort}}]{bustamante:43}
\bibinfo{author}{\bibfnamefont{C.}~\bibnamefont{Bustamante}},
  \bibinfo{author}{\bibfnamefont{J.}~\bibnamefont{Liphardt}}, \bibnamefont{and}
  \bibinfo{author}{\bibfnamefont{F.}~\bibnamefont{Ritort}},
  \bibinfo{journal}{Physics Today} \textbf{\bibinfo{volume}{58}},
  \bibinfo{pages}{43} (\bibinfo{year}{2005}),
  \urlprefix\url{http://link.aip.org/link/?PTO/58/43/1}.

\bibitem[{\citenamefont{Lieb}(1999)}]{Lieb1999491}
\bibinfo{author}{\bibfnamefont{E.~H.} \bibnamefont{Lieb}},
  \bibinfo{journal}{Physica A: Statistical Mechanics and its Applications}
  \textbf{\bibinfo{volume}{263}}, \bibinfo{pages}{491 } (\bibinfo{year}{1999}),
  ISSN \bibinfo{issn}{0378-4371}, \bibinfo{note}{proceedings of the 20th IUPAP
  International Conference on Statistical Physics},
  \urlprefix\url{http://www.sciencedirect.com/science/article/B6TVG-3YGKSSR-1R%
/2/4deb7bfa0074a6a1930a8fc21deed508}.

\bibitem[{\citenamefont{Komatsu et~al.}(2008)\citenamefont{Komatsu, Nakagawa,
  Sasa, and Tasaki}}]{PhysRevLett.100.230602}
\bibinfo{author}{\bibfnamefont{T.~S.} \bibnamefont{Komatsu}},
  \bibinfo{author}{\bibfnamefont{N.}~\bibnamefont{Nakagawa}},
  \bibinfo{author}{\bibfnamefont{S.-i.} \bibnamefont{Sasa}}, \bibnamefont{and}
  \bibinfo{author}{\bibfnamefont{H.}~\bibnamefont{Tasaki}},
  \bibinfo{journal}{Phys. Rev. Lett.} \textbf{\bibinfo{volume}{100}},
  \bibinfo{pages}{230602} (\bibinfo{year}{2008}).

\bibitem[{\citenamefont{Cercignani}(1988)}]{cercignani1988}
\bibinfo{author}{\bibfnamefont{C.}~\bibnamefont{Cercignani}},
  \emph{\bibinfo{title}{The Boltzmann equation and its applications}}
  (\bibinfo{publisher}{Springer}, \bibinfo{year}{1988}).

\bibitem[{\citenamefont{Zakharov et~al.}(1992)\citenamefont{Zakharov, L'vov,
  and Falkovich}}]{zakharov41kst}
\bibinfo{author}{\bibfnamefont{V.}~\bibnamefont{Zakharov}},
  \bibinfo{author}{\bibfnamefont{V.}~\bibnamefont{L'vov}}, \bibnamefont{and}
  \bibinfo{author}{\bibnamefont{Falkovich}}, \emph{\bibinfo{title}{{Kolmogorov
  Spectra of Turbulence 1: Wave Turbulence}}}
  (\bibinfo{publisher}{Springer-Verlag}, \bibinfo{year}{1992}).

\bibitem[{\citenamefont{Kolmogorov}(1941)}]{kolmogorov41}
\bibinfo{author}{\bibfnamefont{A.}~\bibnamefont{Kolmogorov}}, in
  \emph{\bibinfo{booktitle}{Proceedings (Doklady) Academy of Sciences, USSR}}
  (\bibinfo{year}{1941}), vol.~\bibinfo{volume}{30}, pp.
  \bibinfo{pages}{301--305}.

\bibitem[{\citenamefont{Frisch}(1995)}]{frisch1995t}
\bibinfo{author}{\bibfnamefont{U.}~\bibnamefont{Frisch}},
  \emph{\bibinfo{title}{{Turbulence: the legacy of AN Kolmogorov}}}
  (\bibinfo{publisher}{Cambridge University Press}, \bibinfo{year}{1995}).

\bibitem[{\citenamefont{Janssen}(2004)}]{janssen2004iow}
\bibinfo{author}{\bibfnamefont{P.}~\bibnamefont{Janssen}},
  \emph{\bibinfo{title}{{The Interaction of Ocean Waves and Wind}}}
  (\bibinfo{publisher}{Cambridge University Press}, \bibinfo{year}{2004}).

\bibitem[{\citenamefont{Onorato et~al.}(2002)\citenamefont{Onorato, Osborne,
  Serio, Resio, Pushkarev, Zakharov, and Brandini}}]{onorato2002fdw}
\bibinfo{author}{\bibfnamefont{M.}~\bibnamefont{Onorato}},
  \bibinfo{author}{\bibfnamefont{A.}~\bibnamefont{Osborne}},
  \bibinfo{author}{\bibfnamefont{M.}~\bibnamefont{Serio}},
  \bibinfo{author}{\bibfnamefont{D.}~\bibnamefont{Resio}},
  \bibinfo{author}{\bibfnamefont{A.}~\bibnamefont{Pushkarev}},
  \bibinfo{author}{\bibfnamefont{V.}~\bibnamefont{Zakharov}}, \bibnamefont{and}
  \bibinfo{author}{\bibfnamefont{C.}~\bibnamefont{Brandini}},
  \bibinfo{journal}{Physical Review Letters} \textbf{\bibinfo{volume}{89}},
  \bibinfo{pages}{144501} (\bibinfo{year}{2002}).

\bibitem[{\citenamefont{Dyachenko et~al.}(2003)\citenamefont{Dyachenko,
  Korotkevich, and Zakharov}}]{dyachenko2003wtg}
\bibinfo{author}{\bibfnamefont{A.}~\bibnamefont{Dyachenko}},
  \bibinfo{author}{\bibfnamefont{A.}~\bibnamefont{Korotkevich}},
  \bibnamefont{and} \bibinfo{author}{\bibfnamefont{V.}~\bibnamefont{Zakharov}},
  \bibinfo{journal}{Journal of Experimental and Theoretical Physics Letters}
  \textbf{\bibinfo{volume}{77}}, \bibinfo{pages}{546} (\bibinfo{year}{2003}).

\bibitem[{\citenamefont{Lvov et~al.}(2004)\citenamefont{Lvov, Polzin, and
  Tabak}}]{lvov2004eso}
\bibinfo{author}{\bibfnamefont{Y.}~\bibnamefont{Lvov}},
  \bibinfo{author}{\bibfnamefont{K.}~\bibnamefont{Polzin}}, \bibnamefont{and}
  \bibinfo{author}{\bibfnamefont{E.}~\bibnamefont{Tabak}},
  \bibinfo{journal}{Physical Review Letters} \textbf{\bibinfo{volume}{92}},
  \bibinfo{pages}{128501} (\bibinfo{year}{2004}).

\bibitem[{\citenamefont{Dyachenko et~al.}(1992)\citenamefont{Dyachenko, Newell,
  Pushkarev, and Zakharov}}]{dyachenko:1992hc}
\bibinfo{author}{\bibfnamefont{S.}~\bibnamefont{Dyachenko}},
  \bibinfo{author}{\bibfnamefont{A.~C.} \bibnamefont{Newell}},
  \bibinfo{author}{\bibfnamefont{A.}~\bibnamefont{Pushkarev}},
  \bibnamefont{and} \bibinfo{author}{\bibfnamefont{V.~E.}
  \bibnamefont{Zakharov}}, \bibinfo{journal}{Physica D: Nonlinear Phenomena}
  \textbf{\bibinfo{volume}{57}}, \bibinfo{pages}{96} (\bibinfo{year}{1992}),
  \urlprefix\url{http://www.sciencedirect.com/science/article/B6TVK-46JH21H-4G%
/2/b9bf3a47086f6f154a8c0478ca64c07b}.

\bibitem[{\citenamefont{Berloff and Svistunov}(2002)}]{berloff2002ssn}
\bibinfo{author}{\bibfnamefont{N.}~\bibnamefont{Berloff}} \bibnamefont{and}
  \bibinfo{author}{\bibfnamefont{B.}~\bibnamefont{Svistunov}},
  \bibinfo{journal}{Physical Review A} \textbf{\bibinfo{volume}{66}},
  \bibinfo{pages}{13603} (\bibinfo{year}{2002}).

\bibitem[{\citenamefont{Nazarenko and Onorato}(2006)}]{nazarenko2006wta}
\bibinfo{author}{\bibfnamefont{S.}~\bibnamefont{Nazarenko}} \bibnamefont{and}
  \bibinfo{author}{\bibfnamefont{M.}~\bibnamefont{Onorato}},
  \bibinfo{journal}{Physica D: Nonlinear Phenomena}
  \textbf{\bibinfo{volume}{219}}, \bibinfo{pages}{1} (\bibinfo{year}{2006}).

\bibitem[{\citenamefont{Proment et~al.}(2009)\citenamefont{Proment, Nazarenko,
  and Onorato}}]{proment:051603}
\bibinfo{author}{\bibfnamefont{D.}~\bibnamefont{Proment}},
  \bibinfo{author}{\bibfnamefont{S.}~\bibnamefont{Nazarenko}},
  \bibnamefont{and} \bibinfo{author}{\bibfnamefont{M.}~\bibnamefont{Onorato}},
  \bibinfo{journal}{Physical Review A (Atomic, Molecular, and Optical Physics)}
  \textbf{\bibinfo{volume}{80}}, \bibinfo{eid}{051603}
  (pages~\bibinfo{numpages}{4}) (\bibinfo{year}{2009}),
  \urlprefix\url{http://link.aps.org/abstract/PRA/v80/e051603}.

\bibitem[{\citenamefont{Galtier et~al.}(2000)\citenamefont{Galtier, Nazarenko,
  Newell, and Pouquet}}]{galtier2000wtt}
\bibinfo{author}{\bibfnamefont{S.}~\bibnamefont{Galtier}},
  \bibinfo{author}{\bibfnamefont{S.}~\bibnamefont{Nazarenko}},
  \bibinfo{author}{\bibfnamefont{A.}~\bibnamefont{Newell}}, \bibnamefont{and}
  \bibinfo{author}{\bibfnamefont{A.}~\bibnamefont{Pouquet}},
  \bibinfo{journal}{Journal of Plasma Physics} \textbf{\bibinfo{volume}{63}},
  \bibinfo{pages}{447} (\bibinfo{year}{2000}).

\bibitem[{\citenamefont{Kats et~al.}(1975)\citenamefont{Kats, Kontorovich,
  Moiseev, and Novikov}}]{kats1975}
\bibinfo{author}{\bibfnamefont{A.}~\bibnamefont{Kats}},
  \bibinfo{author}{\bibfnamefont{V.}~\bibnamefont{Kontorovich}},
  \bibinfo{author}{\bibfnamefont{S.}~\bibnamefont{Moiseev}}, \bibnamefont{and}
  \bibinfo{author}{\bibfnamefont{V.}~\bibnamefont{Novikov}},
  \bibinfo{journal}{ZhETF Pis ma Redaktsiiu} \textbf{\bibinfo{volume}{21}},
  \bibinfo{pages}{13} (\bibinfo{year}{1975}).

\bibitem[{\citenamefont{Kats}(1976)}]{kats1976}
\bibinfo{author}{\bibfnamefont{A.}~\bibnamefont{Kats}},
  \bibinfo{journal}{Soviet Journal of Experimental and Theoretical Physics}
  \textbf{\bibinfo{volume}{44}}, \bibinfo{pages}{1106} (\bibinfo{year}{1976}).

\bibitem[{\citenamefont{Connaughton and Nazarenko}(2004)}]{nazarenko2004warm}
\bibinfo{author}{\bibfnamefont{C.}~\bibnamefont{Connaughton}} \bibnamefont{and}
  \bibinfo{author}{\bibfnamefont{S.}~\bibnamefont{Nazarenko}},
  \bibinfo{journal}{Phys. Rev. Lett.} \textbf{\bibinfo{volume}{92}},
  \bibinfo{pages}{044501} (\bibinfo{year}{2004}).

\bibitem[{\citenamefont{Karas et~al.}(1976)\citenamefont{Karas, Moiseev, and
  Novikov}}]{karas1976}
\bibinfo{author}{\bibfnamefont{V.}~\bibnamefont{Karas}},
  \bibinfo{author}{\bibfnamefont{S.}~\bibnamefont{Moiseev}}, \bibnamefont{and}
  \bibinfo{author}{\bibfnamefont{V.}~\bibnamefont{Novikov}},
  \bibinfo{journal}{Zhurnal Eksperimental'noi i Teoreticheskoi Fiziki}
  \textbf{\bibinfo{volume}{71}}, \bibinfo{pages}{1421} (\bibinfo{year}{1976}).

\bibitem[{\citenamefont{Connaughton et~al.}(2003)\citenamefont{Connaughton,
  Nazarenko, and Newell}}]{connaughton:2003lg}
\bibinfo{author}{\bibfnamefont{C.}~\bibnamefont{Connaughton}},
  \bibinfo{author}{\bibfnamefont{S.}~\bibnamefont{Nazarenko}},
  \bibnamefont{and} \bibinfo{author}{\bibfnamefont{A.~C.}
  \bibnamefont{Newell}}, \bibinfo{journal}{Physica D: Nonlinear Phenomena}
  \textbf{\bibinfo{volume}{184}}, \bibinfo{pages}{86} (\bibinfo{year}{2003}),
  \urlprefix\url{http://www.sciencedirect.com/science/article/B6TVK-49CRR6J-1/%
2/278d7b4065f6440d2117abc08ee9ac48}.

\bibitem[{\citenamefont{Balk}(2000)}]{balk2000kzs}
\bibinfo{author}{\bibfnamefont{A.}~\bibnamefont{Balk}},
  \bibinfo{journal}{Physica D: Nonlinear Phenomena}
  \textbf{\bibinfo{volume}{139}}, \bibinfo{pages}{137} (\bibinfo{year}{2000}).

\bibitem[{\citenamefont{Connaughton}(2009)}]{connaughton2009ns}
\bibinfo{author}{\bibfnamefont{C.}~\bibnamefont{Connaughton}},
  \bibinfo{journal}{Physica D: Nonlinear Phenomena}
  \textbf{\bibinfo{volume}{238}}, \bibinfo{pages}{2282 }
  (\bibinfo{year}{2009}), ISSN \bibinfo{issn}{0167-2789},
  \urlprefix\url{http://www.sciencedirect.com/science/article/B6TVK-4X85F57-1/%
2/e0db658e124585ac3a87437017414d85}.

\bibitem[{\citenamefont{Hasselmann et~al.}(1985)\citenamefont{Hasselmann,
  Hasselmann, Allender, and Barnett}}]{hasselmann1985c2}
\bibinfo{author}{\bibfnamefont{S.}~\bibnamefont{Hasselmann}},
  \bibinfo{author}{\bibfnamefont{K.}~\bibnamefont{Hasselmann}},
  \bibinfo{author}{\bibfnamefont{J.}~\bibnamefont{Allender}}, \bibnamefont{and}
  \bibinfo{author}{\bibfnamefont{T.}~\bibnamefont{Barnett}},
  \bibinfo{journal}{J. Phys. Oceanogr} \textbf{\bibinfo{volume}{15}},
  \bibinfo{pages}{1378} (\bibinfo{year}{1985}).

\bibitem[{\citenamefont{Leith}(1967)}]{leith1967}
\bibinfo{author}{\bibfnamefont{C.~E.} \bibnamefont{Leith}},
  \bibinfo{journal}{Physics of Fluids} \textbf{\bibinfo{volume}{10}},
  \bibinfo{pages}{1409} (\bibinfo{year}{1967}),
  \urlprefix\url{http://link.aip.org/link/?PFL/10/1409/1}.

\bibitem[{\citenamefont{L'vov and Nazarenko}(2006)}]{lvov2006di}
\bibinfo{author}{\bibfnamefont{V.}~\bibnamefont{L'vov}} \bibnamefont{and}
  \bibinfo{author}{\bibfnamefont{S.}~\bibnamefont{Nazarenko}},
  \bibinfo{journal}{JETP letters} \textbf{\bibinfo{volume}{83}},
  \bibinfo{pages}{541} (\bibinfo{year}{2006}).

\bibitem[{\citenamefont{Boffetta et~al.}(2009)\citenamefont{Boffetta, Celani,
  Dezzani, Laurie, and Nazarenko}}]{boffetta2009mo}
\bibinfo{author}{\bibfnamefont{G.}~\bibnamefont{Boffetta}},
  \bibinfo{author}{\bibfnamefont{A.}~\bibnamefont{Celani}},
  \bibinfo{author}{\bibfnamefont{D.}~\bibnamefont{Dezzani}},
  \bibinfo{author}{\bibfnamefont{J.}~\bibnamefont{Laurie}}, \bibnamefont{and}
  \bibinfo{author}{\bibfnamefont{S.}~\bibnamefont{Nazarenko}},
  \bibinfo{journal}{Journal of Low Temperature Physics}
  \textbf{\bibinfo{volume}{156}}, \bibinfo{pages}{193} (\bibinfo{year}{2009}).

\bibitem[{\citenamefont{Peacock}(1999)}]{peacock1999}
\bibinfo{author}{\bibfnamefont{J.}~\bibnamefont{Peacock}},
  \emph{\bibinfo{title}{Cosmological physics}} (\bibinfo{publisher}{Cambridge
  Univ Pr}, \bibinfo{year}{1999}).

\bibitem[{\citenamefont{Lvov et~al.}(1998)\citenamefont{Lvov, Binder, and
  Newell}}]{lvov1998quantum}
\bibinfo{author}{\bibfnamefont{Y.}~\bibnamefont{Lvov}},
  \bibinfo{author}{\bibfnamefont{R.}~\bibnamefont{Binder}}, \bibnamefont{and}
  \bibinfo{author}{\bibfnamefont{A.}~\bibnamefont{Newell}},
  \bibinfo{journal}{Physica D: Nonlinear Phenomena}
  \textbf{\bibinfo{volume}{121}}, \bibinfo{pages}{317} (\bibinfo{year}{1998}),
  ISSN \bibinfo{issn}{0167-2789}.

\bibitem[{\citenamefont{Zakharov and Pushkarev}(1999)}]{zakharov1999diffusion}
\bibinfo{author}{\bibfnamefont{V.}~\bibnamefont{Zakharov}} \bibnamefont{and}
  \bibinfo{author}{\bibfnamefont{A.}~\bibnamefont{Pushkarev}},
  \bibinfo{journal}{Nonlinear Processes in Geophysics}
  \textbf{\bibinfo{volume}{6}}, \bibinfo{pages}{1} (\bibinfo{year}{1999}).

\bibitem[{\citenamefont{Asinari}(2010)}]{Asinari20101776}
\bibinfo{author}{\bibfnamefont{P.}~\bibnamefont{Asinari}},
  \bibinfo{journal}{Computer Physics Communications}
  \textbf{\bibinfo{volume}{181}}, \bibinfo{pages}{1776 }
  (\bibinfo{year}{2010}), ISSN \bibinfo{issn}{0010-4655},
  \urlprefix\url{http://www.sciencedirect.com/science/article/B6TJ5-50HP2XY-4/%
2/31b9575c0bbef1cba9ce59798563cc5f}.

\bibitem[{\citenamefont{Newell et~al.}(2001)\citenamefont{Newell, Nazarenko,
  and Biven}}]{Newell2001520}
\bibinfo{author}{\bibfnamefont{A.~C.} \bibnamefont{Newell}},
  \bibinfo{author}{\bibfnamefont{S.}~\bibnamefont{Nazarenko}},
  \bibnamefont{and} \bibinfo{author}{\bibfnamefont{L.}~\bibnamefont{Biven}},
  \bibinfo{journal}{Physica D: Nonlinear Phenomena}
  \textbf{\bibinfo{volume}{152-153}}, \bibinfo{pages}{520 }
  (\bibinfo{year}{2001}), ISSN \bibinfo{issn}{0167-2789},
  \urlprefix\url{http://www.sciencedirect.com/science/article/B6TVK-430G97Y-1V%
/2/7539b751b3e972419ca247108cd8d56c}.

\end{thebibliography}

\end{document}